\documentclass{ws-ijmpb}

\begin{document}

\input epsf.sty

\markboth{B. V. Fine}
{LONG-TIME RELAXATION ON SPIN LATTICE AS 
A MANIFESTATION OF CHAOTIC DYNAMICS}

\catchline{}{}{}{}{}
%
%%%%%%%%%%%%%%%%%%%%%%%%%%%%%%%%%%%%%%%%%%%%%%%%%%%%%%%%%%%%%%%%%%%%

\title{LONG-TIME RELAXATION ON SPIN LATTICE AS 
A MANIFESTATION OF CHAOTIC DYNAMICS  }

\author{BORIS V. FINE}

\address{Max Planck Institute for the Physics of Complex Systems, 
Noethnitzer Str. 38\\
01187 Dresden, Germany\\
fine@mpipks-dresden.mpg.de}

\maketitle

\begin{history}
\received{Day Month Year}
\revised{Day Month Year}
%\accepted{(Day Month Year)}
%\comby{(xxxxxxxxxx)}
\end{history}

\begin{abstract}

The long-time behavior of the infinite temperature 
spin correlation functions 
describing the free induction decay in nuclear magnetic resonance and
intermediate structure factors in inelastic neutron scattering is considered.
These correlation functions are defined for  one-, two- and
three-dimensional infinite lattices of interacting spins both
classical and quantum. It is shown
that, even though the
characteristic timescale of the long-time decay of the correlation functions
considered is non-Markovian, the generic functional form of 
this decay is either simple
exponential or exponential multiplied by cosine.
This work contains (i) summary of the existing experimental
and numerical evidence of the above asymptotic behavior;  
(ii) theoretical explanation of this behavior;  and 
(iii) semi-empirical analysis of various 
factors discriminating between the monotonic and the oscillatory long-time
decays.
The theory is
based on a fairly strong conjecture that, as a result of chaos generated by the
spin dynamics, a Brownian-like Markovian description can be applied to the
long-time properties of ensemble average quantities on a non-Markovian
timescale.  The formalism resulting from that conjecture can be
described as ``correlated diffusion in finite volumes.'' 

\end{abstract}


\section{Introduction}
\label{intro}

One of the theoretical challenges associated with strongly interacting
many-body systems is how
to calculate the fast relaxation that takes place on a timescale  of the
order of the dynamic memory time of individual particles.  Generic problems
of this kind  do not have enough small parameters to be solved in a 
controllable approximation. In such a situation, it is  the
intuitive concept of chaotic motion that, nevertheless, preserves the
hope that relatively simple theories can make
quantitative predictions on the fast timescale.  An important question
in this context is:  If the system is, indeed, chaotic, does it lead to
the properties of the fast relaxation, which are independent of the specific
form of the inter-particle interaction? 

In this work\cite{history}, we consider systems of interacting spins---both
classical and quantum---and focus on the functional form of the
long-time behavior of certain fast decaying correlation functions. We
point to experimental and numerical evidence 
that the functional form
in question is surprisingly simple in both classical and spin 1/2 systems.
We then produce an extensive argument asserting that even when conventional
Markovian assumptions are not supposed to apply, it is still possible
to justify quite an unusual diffusion description of the spin systems
based on the assumption of chaotic dynamical mixing.
In the framework of such a description, 
the observed functional form of the  long-time behavior 
appears as a generic property of the correlation functions considered.

\subsection{Quantity of interest}
\label{formulation}

We study the infinite temperature correlation
functions defined on an infinite spin lattice as:  
\begin{equation} 
G(t)= \left\langle \;  S_0^\mu(t) \> \; 
      \hbox{$\sum_{n}$cos({\boldmath $q \cdot r$}$_n$)}
				   \;  S_n^\mu(0) \; \right\rangle,
\label{G} 
\end{equation} 
where $S_n^\mu$ is either the classical projection or
the quantum spin operator
representing 
the $\mu$th ($x$,$y$ or $z$) spin component on the $n$th lattice site;
{\boldmath $r$}$_n$ is the translation vector between the zeroth and
the $n$th sites;
and {\boldmath $q$} is a wave vector corresponding to a spatial period  
commensurate with the lattice periodicity. 
The lattice can have any
Bravais structure in one, two or three dimensions. 
Each spin on the lattice interacts with a finite number of
neighbors according to the Hamiltonian  
\begin{equation} 
{\cal H} =
\sum_{k<n} [J_{kn}^x S_k^x S_n^x + J_{kn}^y S_k^y S_n^y + 
J_{kn}^z S_k^z S_n^z], 
\label{H} 
\end{equation} 
where $J_{kn}^\mu$  
are coupling constants. These constants must be such that  
Hamiltonian (\ref{H}) is 
invariant with respect to lattice translations.
Formula (\ref{G}) distinguishes the zeroth spin from other spins 
only for the convenience of the later discussion.  We shall also use
variable $S$ with no indices to refer to the absolute values
of individual spins.

Given the Hamiltonian (\ref{H}), the timescale of individual spin motion, 
referred to below as ``fast,'' ``short,'' or ``mean free time,'' is
well-represented by the time $\tau$ defined as
\begin{equation}
\tau = \left[ S^2 \sum_{n,\mu}{J_{kn}^\mu}^2 \right]^{-1/2}. 
\label{tau}
\end{equation}
Unless special reasons exist,
this timescale characterizes both the short-
and the long-time decay of  $G(t)$.

Another fact worth mentioning is
that, as a result of the time reversibility of the spin dynamics,
$G(t)$ is an even function of $t$, which, in particular, implies
\mbox{${dG \over dt}|_{t=0}=0$}.

In the context of inelastic neutron scattering, the  correlation
functions (\ref{G}) are referred to as intermediate structure factors
\cite{neutron}.  If $q=0$, Eq.(\ref{G}) can also represent the free
induction decay in nuclear magnetic resonance
(NMR)\cite{LN},\cite{Abragam}.  From the viewpoint of practical
implications, the subject of this work is  more relevant to NMR,
because NMR deals
almost exclusively with high temperatures (on the scale of nuclear
spin energies).

\subsection{Nonequilibrium interpretation}
\label{noneq}

The infinite-temperature limit offers a unique advantage, which will be
important for our subsequent treatment. Namely, 
the correlation function $G(t)$ can be considered as proportional
to the polarization of the zeroth spin 
$\langle S_0^\mu(t) \rangle_{\rho(0)}$ averaged 
with the initial distribution function (or quantum density matrix)
\begin{equation}
\rho(0) \simeq \hbox{exp} [ \; \beta_0 \; 
                   \hbox{$\sum_{n}$cos({\boldmath $q \cdot r$}$_n$)}
				   \;  S_n^\mu(0) \; ],
\label{rho}
\end{equation}
provided the  inverse temperature $\beta_0$ is very small, and only the 
first order of  $\beta_0$ is kept in the expansion of 
$\langle S_0^\mu(t) \rangle$.
In other words, we are dealing with the decay of the
average spin polarization on the zeroth lattice site, given
{\it weak} \ initial polarization on each site proportional
to cos({\boldmath $q \cdot r$}$_n$). 
For the sake of physical interpretation, one can imagine that, at $t<0$,
the Hamiltonian (\ref{H}) was switched off, and 
the spin system was equilibrated at very high temperature 
in an external field 
varying as cos({\boldmath $q \cdot r$}$_n$). Then, at $t=0$, the
environment and the 
external field were switched off. Simultaneously, 
the Hamiltonian (\ref{H}) was switched on,
and the  system started evolving
to a new equilibrium, which, to the first order in $\beta_0$,
corresponds to
infinite temperature.
This interpretation is visualized in Fig.~\ref{examples} 
for classical spins.

\setlength{\unitlength}{0.1cm}
\begin{figure}
\begin{picture}(82,0)
\end{picture}
\begin{picture}(82,25)
{
\put(10,5){

\epsfxsize=2.5in
\epsfbox{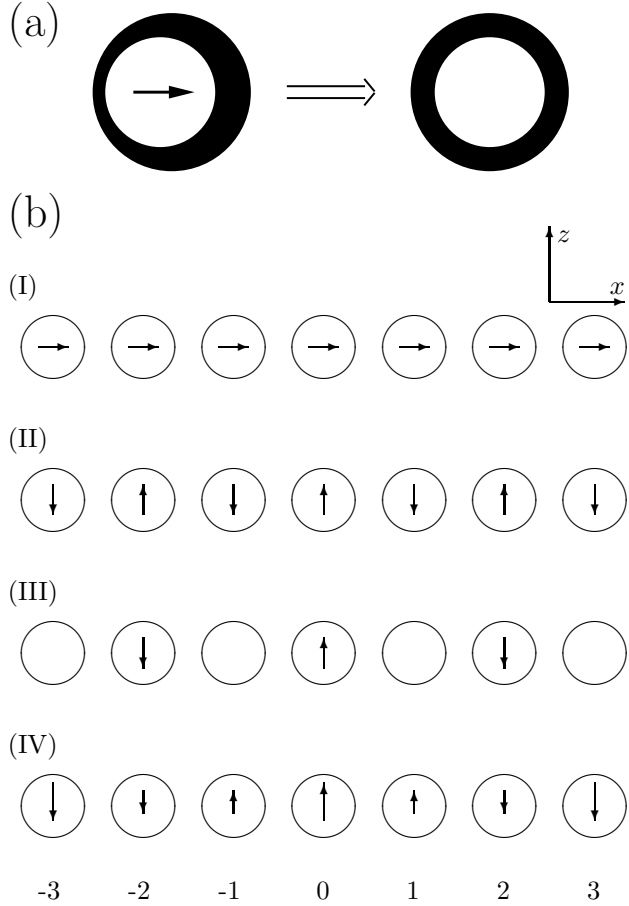} }
}
\put(0,23){\LARGE (a)}
\end{picture}
\begin{picture}(82,30)

\put(72,18){\vector(1,0){10}}
\put(72,18){\vector(0,1){10}}
\put(73,26){$z$}
\put(80,19){$x$}
\put(0,19){(I)}
\put(0,28){\LARGE (b)}
\multiput(4,12)(12,0){7}{\vector(1,0){4}}
\multiput(6,12)(12,0){7}{\circle{8}}

\end{picture}
\begin{picture}(82,20)

\put(0,19){(II)}
\multiput(6,14)(24,0){4}{\vector(0,-1){4}}
\multiput(18,10)(24,0){3}{\vector(0,1){4}}
\multiput(6,12)(12,0){7}{\circle{8}}
\end{picture}
\begin{picture}(82,20)

\put(0,19){(III)}
\multiput(18,14)(48,0){2}{\vector(0,-1){4}}
\put(42,10){\vector(0,1){4}}
\multiput(6,12)(12,0){7}{\circle{8}}

\end{picture}
\begin{picture}(82,20)

\put(0,19){(IV)}
\multiput(6,15)(72,0){2}{\vector(0,-1){5}}
\put(18,14){\vector(0,-1){3}}
\put(30,11){\vector(0,1){3}}
\put(42,10){\vector(0,1){5}}
\put(54,11){\vector(0,1){3}}
\put(66,14){\vector(0,-1){3}}
\multiput(6,12)(12,0){7}{\circle{8}}

\end{picture}
\begin{picture}(82,5)

\put(4,5){-3}
\put(16,5){-2}
\put(28,5){-1}
\put(41,5){0}
\put(53,5){1}
\put(65,5){2}
\put(77,5){3}

\end{picture}
\caption{
These pictures visualize the nonequilibrium problem
corresponding to the initial probability distribution
(\protect{\ref{rho}}) for  classical spins.
Picture (a) represents an example of
the initial and the final probability distributions
of the zeroth classical spin.
The tip of the spin vector 
moves on a spherical surface 
shown as a white disk.  
The thickness of the outer black layer around that disc
represents the probability to find the spin  oriented
in the corresponding radial direction. 
The arrow in the middle of the white
disc indicates  
the average spin polarization.
Thus the problem is one of calculating the
time dependence of the average spin polarization,
given that the probability density on the spherical surface 
evolves from  a weakly anisotropic distribution (left) to the
completely isotropic infinite temperature distribution (right).
Picture (b) shows examples of the initial conditions for  
fragments of infinite spin chains. The arrows indicate the direction
and the relative size of the {\it weak} spin polarization at $t=0$. 
The
numbers in the bottom line are the indices of the spin sites.
Examples (I-IV)  correspond to the calculation of
the correlation functions of form 
(\protect{\ref{G}}), provided 
(I) $\mu \to x$, $q = 0$;
(II) $\mu \to z$, $q = \pi$;
(III) $\mu \to z$, $q = \pi/2$;
(IV) $\mu \to z$, $q = \pi/3$.
} 
\label{examples}
\end{figure}

\subsection{Theoretical goal}
\label{goal}

In this work, we intend to show theoretically 
that generic long-time behavior of $G(t)$
has one of the following two
functional forms: either
\begin{equation} 
G(t) \simeq e^{- \xi t},
\label{longexp} 
\end{equation} 
or 
\begin{equation} 
G(t) \simeq e^{- \xi t} cos(\eta t + \phi), 
\label{longcos} 
\end{equation} 
where $\xi$, $\eta$ and $\phi$ are some constants  about which we can
only assert that, in general, the values of $\xi$ and
$\eta$ are of the order of $1/\tau$.
It will follow from both the empirical evidence and our theory that
$G(t)$ approaches the asymptotic form (\ref{longexp}) or 
(\ref{longcos}) after a time of the order of several $\tau$,
i.e. sufficiently fast.

We define the  ``generic long-time behavior'' as follows:

For a given lattice and for a given radius of interaction,
there is a finite number of independent interaction
coefficients necessary to specify the translationally
invariant Hamiltonian (\ref{H}).
(For example, for the spin chain  with 
the nearest neighbor interaction of form(\ref{H}),
one must specify only three coefficients $J^x$, $J^y$ and $J^z$.)
If, we  restrict the value of each independent interaction
coefficient to a finite interval (the same for all of them) 
and, from that interval,  pick
the values of those coefficients randomly,
then our claim is that, with probability 1,
the long-time behavior of the correlation functions (\ref{G})
will have the functional form either 
(\ref{longexp}) or (\ref{longcos}).

The above definition implies that there can be infinitely many exceptions
not exhibiting the long-time behavior (\ref{longexp}, \ref{longcos}) --- 
 we cannot name
all of them --- but the overwhelmingly general rule is still given by
Eqs.(\ref{longexp}, \ref{longcos}).

\

It is important to realize that, if the functional dependence
(\ref{longexp}, \ref{longcos}) is, indeed, generic, then
this property is very likely  related to the
randomness generated by the spin dynamics. At the same time,
the problem cannot be reduced to
the Markovian paradigm of
``a slow variable interacting with a fast equilibrating
background'' --- the decay (\ref{longexp}) or 
(\ref{longcos}) occurs on the timescale of $\tau$, 
which is the fastest natural timescale of the problem.
There would be no contradiction to the standard theory
of Brownian-type motion (which is based on the above
paradigm) if, on the timescale of $\tau$,
the long-time decay of correlation functions (\ref{G})
was described by a power law, Gaussian, or by  some
erratic functional form --- different for different correlation functions.
Therefore, whatever is the ultimate explanation of 
the long-time behavior (\ref{longexp},\ref{longcos}),
it will certainly be a step 
beyond the standard theory of Brownian-type motion. 
In particular, 
formulas~(\ref{longexp},\ref{longcos})
cannot be viewed as yet another example of the damped harmonic oscillator,
because the standard treatment of the damped
oscillator requires the separation of timescales between the slow
oscillator and the quickly equilibrating microscopic motion in a heat bath.

It is also worth mentioning that the oscillations in
Eq.(\ref{longcos}) are neither related to a simple rotation nor
produced by the equivalent of a simple spring potential.   
Below, we illustrate the non-trivial nature of this fact 
by considering the example of
the  NMR free induction decay, in which case
the oscillations have actually been observed experimentally 
(see Section~\ref{empirical}). 

The NMR free induction decay is the decay of the macroscopic magnetization
of  nuclear spins under the influence of local fields fluctuating
around the zero average value.  When present, 
the oscillations of this decay are not
the result of any kind of  magnetization precession, i.e.,  at the
moment when the free induction decay crosses zero, the magnetization
of the system is equal to zero {\it in all directions}, and then, out of 
completely unpolarized state the magnetization reappears with the
opposite sign.
Such a fact is hard  to understand intuitively, because,
the value of the macroscopic magnetization appears to be  a
direct measure of the deviation from equilibrium, but,
at the same time,  normal
intuition, which is supported by numerous examples of the Markovian
limit,  would suggest that, once at equilibrium,
the system should not deviate from it (beyond the range
of usual thermal fluctuations). It is then even harder to 
explain how the fluctuations of local fields
cause the oscillations of the long-time decay to be periodic.

\

The importance of the long-time behavior (\ref{longexp},\ref{longcos})
should not be underestimated in light of the fact that there are many
ways (memory functions, continued fractions, etc.) to produce such a
behavior with very simple ingredients. It is precisely the
justification of those ingredients that makes the whole issue
intractable within any of those attempts.

In particular, a number of NMR-related 
theories\cite{Tjon,PL,BW,BW1,EC,Jensen,Lundin,Fine1}
would predict the long-time
behavior given by Eqs.(\ref{longexp},\ref{longcos}).  However, all of
those theories struggle to achieve an effective approximation for the
entire evolution of $G(t)$, of which the long-time part is not the most
prominent one. As a result, the long-time functional dependence becomes
a side effect of quite crude and uncontrollable assumptions adopted
only for the sake of simplicity.  Therefore, the long-time predictions
of those theories remain doubtful.  (See e.g. Ref.~\cite{Cowan}.)
In order to illustrate this point, 
in Appendix~\ref{critique} we  discuss the treatment 
of Borckmans and Walgraef\cite{BW,BW1},
who have made a particularly strong claim of  the derivation
of the long-time functional form (\ref{longexp},~\ref{longcos}).

\

The theory to be presented in this work is  unusual
in one important respect.
Namely, it leads to the definite 
result (\ref{longexp}, \ref{longcos}) for the
functional form of the long-time decay without giving the recipe for
calculating the parameters describing that decay. 
Our viewpoint is that should the validity of
Eqs.(\ref{longexp},\ref{longcos}) be shown in a model-independent way,
the approximate schemes can rely on such a result rather than derive
it. As a consequence, the status of the approximate calculations
would change from 
an uncontrollable extrapolation in the time domain to something more
similar to interpolation. In order to obtain the
necessary long-time parameters, it will be sufficient to compute the
initial evolution of the
correlation functions  up to the point, where they are supposed
to be describable by Eq.(\ref{longexp}) or (\ref{longcos}).
In that scheme, the approximate analytical
expressions for the initial evolution will not be required
to have the asymptotic
long-time form (\ref{longexp}) or (\ref{longcos}).

The existing approximation 
schemes\cite{LN,Tjon,PL,BW,BW1,EC,Jensen,Lundin,Fine1,Cowan,Shakhmuratov} 
are already quite good in describing
the initial behavior of correlation functions, but, as far as
the parameters of the long-time behavior are concerned, 
the {\it a priori} accuracy of {\it predictions} based on those schemes 
is yet to be established. (Our estimate is that it is not better
than 20 per cent.)

Another unusual feature of our theory is that it
relies on a relatively long chain of  qualitative arguments.
Though not a rigorous derivation, such a treatment  
allows us to address the questions,
which were mostly ignored in the previous works.  Namely:

Why is the long-time regime universal? 

What makes that regime different from the initial regime?

Why does it have such a functional form?

Why are Eqs.(\ref{longexp},\ref{longcos}) 
primarily relevant to  the 
$q$-dependent correlation functions (\ref{G})
and not to the pair correlation functions
of the form $\left\langle \;  S_k^\mu(t)  \; 
S_n^\mu(0) \; \right\rangle$?

What mechanism is responsible for the oscillations in Eq.(\ref{longcos})?

The answers to these questions will be
given in Section~\ref{summary}.

\

A comment is now due on our use of term ``chaos.''

What  we  call ``chaos''
is a concept stronger than Boltzmann's ``molecular chaos'',  because
``molecular chaos'' only postulates the irrelevance of  fast
decaying correlations to slow observables, while in this work we
address precisely the issue of how those fast correlations decay.  At
the same time, our assumptions of chaos are formulated mostly in terms
of the motion of a single particle, which is consistent with, but less
restrictive than, the mathematical definition of chaos related to the
exponential instabilities of trajectories in the phase space of the entire
system.  In particular, our treatment will not be altered if 
the assumed single particle properties result from taking the
thermodynamic limit in the systems not quite complying with the
mathematical definition of chaos.

We should also mention that there are apparent 
parallels between our treatment and the mathematical theory
of Policott-Ruelle resonances in classical chaotic
systems\cite{Ruelle,Gaspard}. However, we are not aware of any definite
mathematical result applicable  to spin lattices with time-independent
translationally invariant Hamiltonians, 
in particular, in the
quantum case.
One of the advantages of our approach is that it allows the
quantum case to be addressed directly, i.e. without any reference to
the classical limit.  Recently, Prosen has arrived to the same conclusion 
about the role of the Policott-Ruelle resonances by considering the decay
of correlation functions of kicked Ising chain of spins 
1/2\cite{Prosen,Prosen1}.

\

The plan of the rest of this paper is the following:
In Section~\ref{empirical} we summarize the empirical
evidence of the long-time behavior (\ref{longexp},\ref{longcos}). 
In Section~\ref{theory}
we develop the theory of the long-time relaxation. 
Finally, in Section~\ref{factors}, we discuss various factors,
which discriminate between the monotonic regime  (\ref{longexp})
and the oscillatory regime (\ref{longcos}).

\section{Empirical evidence}
\label{empirical}

Before reviewing the existing experimental and numerical evidence for the
long-time behavior (\ref{longexp}) and (\ref{longcos}), we recall 
that the characteristic decay time of $G(t)$ is
expected to be of the order of $\tau$ given by Eq.(\ref{tau}). 
At the same time, 
the characteristic long-time behavior does not become
pronounced until after several $\tau$,
i.e. until $G(t)$ has fallen to a relatively small value. 
Yet, in order to make 
definite conclusions about the long-time behavior, one has to go
orders of magnitude below the initial value of $G(t)$. This task is
extremely challenging both numerically and experimentally. In
particular, it is certainly beyond the reach of 
the present day inelastic neutron
scattering experiments.

There exist two pieces of evidence unambiguously showing 
the oscillatory long-time
behavior (\ref{longcos}) in  spin 1/2 systems: 
(i) experiments on NMR free
induction decay in CaF$_2$ by Engelsberg and Lowe \cite{EL}, 
and (ii) the results
of numerical diagonalization of spin $1/2$ chains by Fabricius,
U. L\"ow and J. Stolze  \cite{FLS}. 

In the first case, the underlying theoretical problem involves a simple
cubic lattice of spins 1/2 ($^{19}$F nuclei) coupled by a truncated
magnetic dipolar interaction\cite{Abragam}, 
which has the form (\ref{H}) with 
$J_{kn}^z = -2 J_{kn}^x = - 2 J_{kn}^y \simeq 
{(1- 3 \; \hbox{cos}^2 \vartheta) \over 
|\hbox{\boldmath $r$}_k - \hbox{\boldmath $r$}_n|^3}$, where
$\vartheta$ is the angle between 
$(\hbox{\boldmath $r$}_k - \hbox{\boldmath $r$}_n)$ and the $z$-axis. 
The free
induction decay is then described by Eq.(\ref{G}), with  $q=0$ and $\mu
\rightarrow x$. 
In the context of the experiment, the above quantity is proportional to 
the decay of the average spin polarization in the Larmor rotating
reference frame provided the $z$-axis corresponds to
the direction of the strong external magnetic
field. This implies that the values of 
the interaction coefficients can be changed by 
varying the direction of the external
field with respect to the crystal lattice.   

\begin{figure}
\setlength{\unitlength}{0.1cm}
%=======================================================================
\begin{picture}(82, 180)
{
\put(0,-8){
\epsfxsize=5.7in
\epsfbox{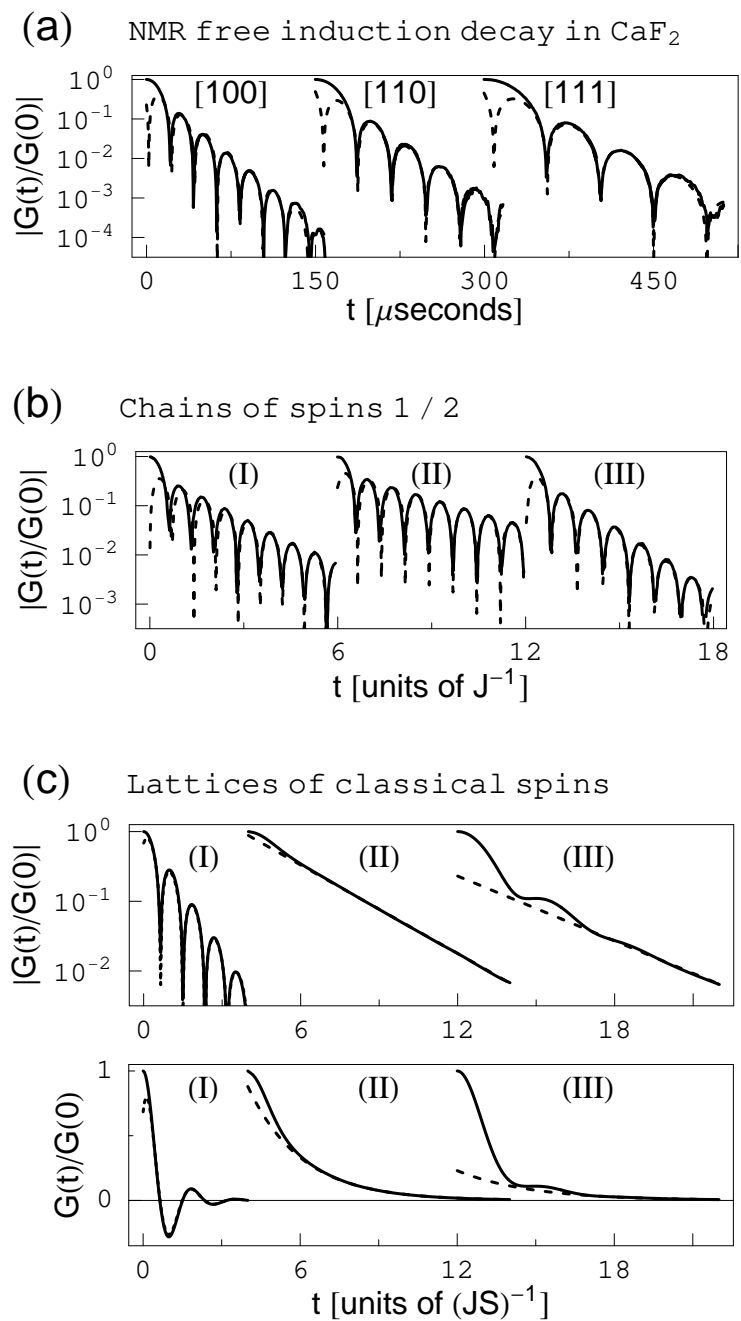} }
}
\end{picture} 
%============== 
\caption{Experimental (a) and numerical (b,c) evidence of the long-time behavior
(\ref{longexp}, \ref{longcos}).
The solid lines represent the data
reproduced from the following sources: (a) Ref.\protect \cite{EL};
(b) Ref.\protect \cite{FLS}; (c) Ref.\protect \cite{Fine3}.
The dashed lines represent long-time fits of the form
(\protect{\ref{longexp}}) or (\protect{\ref{longcos}}).  In each frame,
the origins of the middle and the right plots are displaced along 
the time axis. In all frames, except for the last one, the
absolute values of $G(t)$ are presented on a logarithmic scale,
which leads to cusps when $G(t)$ actually crosses zero.
For reference, the data set (c) set is also presented 
on a direct (non-logarithmic) scale with no absolute value taken.
The details describing each of the plots are given in the text.}
\label{evidence} 
\end{figure}

All three free induction decays
reported in Ref.\cite{EL} are reproduced in Fig.\ref{evidence}(a).
The plots are labelled by the crystal direction 
of the external field. 

It should be mentioned that there is no material other than CaF$_2$, 
where the inter-nuclear
spin-spin interaction is so well-known and so well-isolated from other
microscopic factors, and where, at the same time, the free induction decay is
measured with sufficient accuracy. In a less refined setting, 
the oscillating tails of free induction decays with nodes, 
which, at longer times, become equally spaced 
have also been observed  by Metzger and Gaines\cite{MG} in solidified  
mixtures of molecular hydrogen and deuterium.

The second piece of evidence for the long-time behavior
(\ref{longcos}) comes from the numerical diagonalization of the 
chains (rings) of sixteen
spins 1/2  coupled by the nearest-neighbor interaction of  form
(\ref{H}), with the restriction $J_{n,n+1}^x = J_{n,n+1}^y$. 
These
chains are known as $XXZ$ chains. 
Three  correlation functions of  form
(\ref{G})
have been presented in Ref.\cite{FLS}, and are reproduced in 
Fig.~\ref{evidence}(b).   
Each of those correlation functions
is characterized by the following
parameters entering Eqs.(\ref{G}, \ref{H}): 
$q = \pi$ (in the units of inverse chain spacing), and
$J_{n,n+1}^x = J_{n,n+1}^y = J$, where $J$ is an auxiliary
variable. Other parameters are:
(I) $J_{n,n+1}^z = J$, any value of $\mu$;  
(II) $J_{n,n+1}^z = J \hbox{cos}(0.3 \pi)$, $\mu \to z$;
(III) $J_{n,n+1}^z = J \hbox{cos}(0.3 \pi)$, $\mu \to x$.

In principle, it is an open question to what extent the the spin 1/2
$XXZ$ chains with $J_{n,n+1}^z \neq 0$ display generic properties\cite{FM},
because, on one hand, they are known to be integrable, but, on the
other hand, if attempted, the actual solution is so complex that the
explicit form of the correlation functions of interest has never been
obtained.  As far as the infinite temperature correlation functions are
concerned, the numerical study of those chains\cite{FLS}\cite{FM} did
not reveal anything qualitatively unusual.
Therefore, for now, since not much data on the long-time behavior of
these functions exist, it is reasonable to take the
data on the spin 1/2 $XXZ$ chains as a strong hint for the 
generic nature
of Eq.(\ref{longcos}). However, in the future, if more evidence for the
long-time behavior (\ref{longexp},\ref{longcos}) in generic spin
systems emerges, the logic can be turned around, 
and the existence of such
behavior in the spin 1/2 $XXZ$ chains can be
considered as a piece of evidence proving that the properties of those
integrable chains are, at least, partially generic.

We are not aware of any data 
on  quantum spin systems with time-independent Hamiltonians 
which would
definitely support the monotonic exponential long-time behavior
(\ref{longexp}), when no separation of timescales is present in
the problem.  However, it is  common knowledge in the field of NMR,
that monotonic exponential long-time behavior of free
induction decay occurs even more frequently than oscillatory behavior.
Simple exponential decay of correlation functions 
has also be observed by Prosen in the numerical 
studies of a quantum spin system with time-dependent Hamiltonian
(kicked Ising chain of spins 1/2)\cite{Prosen,Prosen1}.

Monotonic exponential decay is, in a sense, less surprising, given
that at least it is expected  in the situation when the separation of
timescales is present, i.e. when $G(t)$  evolves much slower than a
typical microscopic variable. An
example of such a situation is the problem of ``exchange narrowing''\cite{AW,GHSS},
in which case, the dominating  part of the Hamiltonian (\ref{H}) has
the Heisenberg form ($J_{kn}^x = J_{kn}^y = J_{kn}^z$), and the
correlation function of interest is given by Eq.(\ref{G}) with $q = 0$.
This correlation function does not decay unless the Hamiltonian
contains small corrections to the Heisenberg interaction.
The exponential decay of the slow correlation functions or,
more precisely, the Lorentzian shape of their Fourier transforms has
been verified experimentally\cite{GSJ} in several quantum spin systems
exhibiting exchange narrowing.

Motivated by the spin 1/2 problems and by the theory to be presented below,
we have recently performed numerical
simulations of classical spin systems \cite{Fine3}.  The simulations generated
a wide spectrum of data covering different dimensions, interactions and wave
vectors --- all entirely consistent with the long-time behavior
\mbox{(\ref{longexp}, \ref{longcos})}.   Three of the nine correlation
functions computed in \mbox{Ref.\cite{Fine3}}
are reproduced in Fig.~\ref{evidence}(c). Two of them
demonstrate the monotonic long-time behavior (\ref{longexp}) 
missing in the previous examples. 
Each of the correlation functions presented in Fig.~\ref{evidence}(c)
corresponds to $\mu \to x$ in Eq.(\ref{G}), and
has been computed for  the nearest-neighbor version of Hamiltonian~(\ref{H}).   
The remaining details 
(specific for each plot) are: (I) three-dimensional cubic lattice,
{\boldmath $q$}$= (0,0,0)$, $J_{kn}^x = 0$, $J_{kn}^y = -J$, $J_{kn}^z = J$;
(II) two-dimensional square lattice,
{\boldmath $q$}$= (\pi/2, \pi)$, $J_{kn}^x = 1.2 J$, $J_{kn}^y = -0.2 J$, 
$J_{kn}^z = J$;
(III) one-dimensional chain,
$q = 0$, $J_{kn}^x = 1.2 J$, $J_{kn}^y = -0.3 J$, $J_{kn}^z = J$.

It thus appears that the  long-time behavior represented by
Eqs.(\ref{longexp}, \ref{longcos}) is very real. Its simplicity, 
uncharacteristic of the timescale involved, calls for a
theoretical explanation, which is the subject of the rest of this
work.

\section{Theory of the long-time relaxation}
\label{theory}

\subsection{Outline}
\label{outline}

Our theory is to be developed according to the following plan:  
First, in Section~\ref{cl-intro}, we introduce the classical case but 
then digress and, in Section~\ref{markovian}, analyze one
rarely discussed piece of empirical knowledge related to the Markovian
description. This analysis generates {\it Conjecture I}, on which the
rest of the theory is built.  Section~\ref{markovian}  is quite long, 
but, in fact, reading only the formulation of
{\it Conjecture I} \  is sufficient to proceed to 
Sections~\ref{cl-cont} and  \ref{quantum}.  
In Section~\ref{cl-cont}, we return to the treatment of
classical spins and, in Section~\ref{quantum}, extend that treatment to
quantum spins.
Although the classical limit of the
quantum problem will not be taken, the treatment of quantum spins
will be developed by analogy with 
the classical spins. Therefore, reading the classical section
is essential to understanding of the quantum section. 
Finally, we summarize the theory in Section~\ref{summary}.

\subsection{Classical spins  --- long-time assumption}
\label{cl-intro}

We shall treat the classical case from the viewpoint of the
nonequilibrium interpretation presented in Section~\ref{noneq}
and illustrated in Fig.~\ref{examples}. 
Our attention will be focused on the one-particle 
probability distributions of the
spin orientations.

The evolution of each classical spin vector can be represented as a
trajectory of its tip  on a spherical surface.  Describing the
ensemble of all possible trajectories of a given spin, 
we assume  chaotic mixing of the following kind: 

On the scale characterized by the mean free time $\tau$ given by
Eq.(\ref{tau}) and by the corresponding mean free path 
(of the order of the radius of the sphere), (i)
each trajectory loses the memory of its initial position; (ii) the set
of all possible trajectories starting from any
arbitrary small surface element  disperses over the entire spherical surface
in a random manner; and (iii) the statistics of the trajectory patterns of this 
set
becomes representative of the statistics of the 
whole ensemble of one-spin trajectories. 

The above assumption of chaotic mixing
does not require the statistical properties of every one-spin
trajectory to be representative of the properties of the whole ensemble.
This detail signifies the difference between 
mixing and ergodicity. Following Krylov\cite{Krylov}, we  assume
that  ergodicity is not a necessary condition for  relaxation
to the equilibrium.
The estimate of the mean free time in this
assumption by formula (\ref{tau}) is how the infinite
temperature condition enters our treatment.  At finite temperatures,
the mean free time should be longer, and may even become infinite,
because of the equilibrium correlations between the orientations of
different spins.

If a
point-like particle moves under similar conditions on an infinite
plane, it would be appropriate to apply the theory of Brownian motion
on a scale much greater than the mean free path.  The problem is
that in our case the typical mean free path is of the order of the
size of the sphere.
Therefore, we are interested in an
accurate description on the scale of the order of the mean free path
and smaller.

We now turn to a
general discussion that motivates one additional assumption.

\subsection{Markovian description for the ensembles of trajectories}
\label{markovian}

The standard microscopic theory of Brownian motion is Markovian, i.e.
it employs processes possessing no memory of the past state.  It is
always emphasized (at least in the physics literature)  that such a
description is justified only on a timescale much greater than the mean
free time.

Contrary to the above point, we shall argue 
that, as far as ensemble
averaging is concerned, the adequacy of the
commonly accepted Markovian descriptions typically
propagates to much shorter time scales, and it is only the initial
short-time behavior of a typical theoretical quantity that is not Markovian.

\subsubsection{Exponential decay}
\label{exponential}

The inapplicability of the Markovian description to short (ballistic) 
timescales is
frequently demonstrated by presenting an example of the
nearly exponential decay exhibited by an autocorrelation function having the
general
form 
\begin{equation}
R(t) = \langle X(t) X(0) \rangle.
\label{R}
\end{equation} 
This function characterizes the equilibrium fluctuations of
some
variable $X$ in some macroscopic system (not specified here).
The variable $X$  should evolve much slower
than the rest of the variables describing the same system. In equilibrium,
$\langle X \rangle = 0$, which implies that $R(\infty)=0$.
(An example of such a situation is the exchange narrowing problem 
mentioned in Section~\ref{empirical}.)

The following assumptions with respect to $R(t)$ are usually
appropriate (at least the ``usual wisdom'' suggests so):

{\it Assumption 1:}  The equilibrium correlation function  
$R(t)$ can be equivalently considered as 
characterizing the relaxation of the slow average quantity 
$\langle X(t) \rangle_{\rho_{_R}(0)}$ starting 
from a very small initial value 
$\langle X(0) \rangle_{\rho_{_R}(0)}$ and decaying to the equilibrium zero 
value. 
The notation $\langle \rangle_{\rho_{_R}(0)}$ implies that the averaging 
should be taken over all possible evolutions of the entire system
weighed 
by the probability of the initial conditions  denoted here as
$\rho_{_R}(0)$. The small deviation of  $\rho_{_R}(0)$ from the equilibrium
distribution function should be proportional to $X(0)$.

As explained in Section~\ref{formulation}, as far as the
infinite temperature spin correlation functions are concerned, this kind
of property can be obtained from the high-temperature expansion.
In a general context, the equivalence
between the correlation function of equilibrium  fluctuations 
and the law of
relaxation
was first explicitly postulated by Onsager\cite{Onsager}.

{\it Assumption 2:} 
On the long timescale, the relaxation of the slow quantities can be
represented by Markovian first order rate equations, which can
include only slow variables. These equations have to be linearised
with respect to the small deviations from equilibrium.

{\it Assumption 3:} 
There are no slow variables that can affect $\langle X(t) \rangle$,
other than $\langle X(t) \rangle$ itself.

Given the above assumptions, 
the only Markovian rate equation that can be written
is 
\begin{equation}
{dR \over dt} = - \gamma R,
\label{exp_decay}
\end{equation} 
where the new constant $\gamma$ is the decay rate defined by this equation. 
Assuming $R(0) = 1$, 
the solution of Eq.(\ref{exp_decay}) is 
$R(t) = \hbox{exp}(- \gamma t)$.

It is, usually, at this point that 
the inapplicability of the Markovian description to 
short timescales is illustrated by
pointing out that, on one hand, the Markovian
approximation gives ${dR \over dt}|_{t=0}= - \gamma $, but,
on the other hand, the time reversibility of the underlying dynamics
guarantees that $\langle X(t) X(0) \rangle = \langle X(-t) X(0)
\rangle$,
which means $R(t)$ is an even function of time and, therefore,
${dR \over dt}|_{t=0}=0$.

\begin{figure}
\setlength{\unitlength}{0.1cm}
%=======================================================================
\begin{picture}(82, 65)
{
\put(15,5){

\epsfxsize=3in
\epsfbox{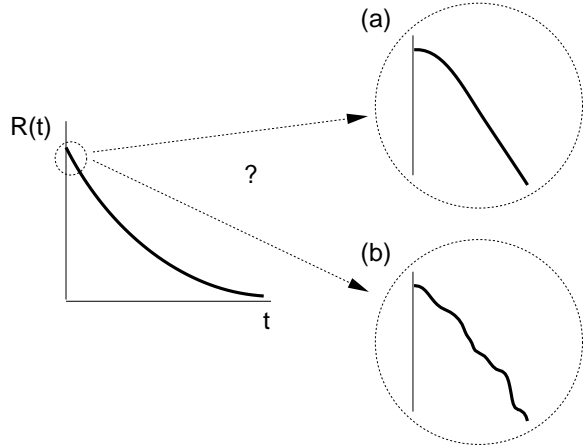} }
}

\end{picture} 
%============== 
\caption[\ Two short-scale pictures of a Markovian correlation function]
{This picture represents the correlation
function $R(t)$ introduced in the text and,
by magnifying the short timescale, shows 
two short-scale pictures compatible with
the long-time Markovian description, namely: (a)
smooth approach to the Markovian behavior with no further
fluctuations  and 
(b) persistent fluctuations around the average Markovian behavior.
It is argued in the text that picture (a) is the one that corresponds to
the reality. } 
\label{magnify} 
\end{figure}

The above contradiction does not invalidate the Markovian description, 
because, {\it a priori},
that description does not claim to provide the exact 
derivatives corresponding to the short-time fluctuations around 
the average long-time behavior.
However, it is  rarely emphasized that, in principle, there are two 
pictures of the short-time behavior of $R(t)$ that can be consistent 
with the long-time Markovian approximation:  
(a)  smooth approach of $R(t)$ to the Markovian
prediction or (b) persistent short-time fluctuations of $R(t)$ around
the average Markovian behavior. 
Both pictures are illustrated in Fig.~\ref{magnify}.

Although, at first sight, it might seem
that picture (b) better corresponds to the spirit of the
Markovian approximation, the fact is:  Whenever a theoretical attempt
is made to embrace both the short- and the long-time behavior of $R(t)$, 
it results in picture (a).  Moreover,
we are not aware of any experiment or numerical calculation 
reliably showing the persistent short-time fluctuations of a correlation
function around the average long-time behavior, provided that
behavior is {\it correctly} predicted from a Markovian description.

Now we illustrate the assertion made in 
the previous paragraph by
showing that the assumption of smooth approach to
Markovian behavior
is the only one that is necessary
in order to obtain 
the Green-Kubo formula \cite{Gaspard,Kubo} 
for the calculation of the constant 
$\gamma$ entering Eq.(\ref{exp_decay}).

The smooth approach to Markovian behavior
shown in Fig.~\ref{magnify}(a) implies that the deviation of
$R(t)$ from the Markovian prediction occurs only around $t=0$, and then
the exact first time derivative of $R(t)$ approaches the
value of $(-\gamma)$ within a short time interval, over which the value
of $R(t)$ itself can change only insignificantly, and, after that,
$R(t)$ follows the Markovian solution without any fluctuations.
Within the same time interval, the second time derivative of $R(t)$
should change dramatically from a certain value characteristic
of the short timescale
to the value complying with the smooth and slow Markovian
behavior. 

The above description
implies the following recipe for the calculation of relaxation rate 
$\gamma$ in Eq.(\ref{exp_decay}): 
(i) Write the exact expression for the second
derivative of $R(t)$  (ii) Perform the time integration
of that expression until the value of the integral
stops changing significantly on the fast timescale of the problem. 
The saturation value of that integral  becomes the first
derivative with which the correlation function enters the Markovian
regime. This value should be equated to $(-\gamma)$.  
In order to present the result in a familiar form,
we note that $\langle \ddot{X}(t) X(0) \rangle$ 
(the exact expression for the second time derivative
of $R(t)$) can be  
transformed\footnote[1]{The 
possibility of such a transformation is the consequence of the fact that 
the origin of the time axis can be arbitrarily shifted by some
value $t_0$ without changing the correlation function of
the equilibrium fluctuations. The transformation proceeds as follows:
\begin{eqnarray}
\nonumber
& \left\langle {d^2X(t) \over dt^2} X(0) \right\rangle  
=  \left\langle {d^2X(t +t_0) \over dt^2} X(t_0) \right\rangle 
\\ \nonumber  
& =  \left\langle {d^2X(t +t_0) \over dt_0^2} X(t_0) \right\rangle  
\\ \nonumber
& = - \left\langle {dX(t +t_0) \over dt_0} {dX(t_0) \over dt_0} \right\rangle
+ {d \over dt_0}\left\langle {dX(t +t_0) \over dt_0} X(t_0)
\right\rangle,
\end{eqnarray}
In equilibrium, the second term obtained above must be
equal to zero, and  the first term is obviously equal to 
$- \langle \dot{X}(t) \dot{X}(0) \rangle$.
} 
to  $-\langle \dot{X}(t) \dot{X}(0) \rangle$.
Therefore,
\begin{equation} 
\gamma =  \int_0^{\infty} \langle \dot{X}(t) \dot{X}(0) \rangle dt. 
\label{G-K}
\end{equation}
The  infinite upper limit of this integral should be understood in an
approximate sense --- it implies an upper cutoff, which is much longer than
the characteristic time of the fast environment but much shorter than
$1/\gamma$. 
In the context of statistical physics, Eq.(\ref{G-K})
is best recognizable  as the Green-Kubo formula, though its resemblance
with the ``golden rule''  of second order perturbation
theory is not a coincidence. 

It is actually easy to find an ansatz of the form
\mbox{$R(t) = \hbox{exp}(- a(t) )$,} where the function $a(t)$
quickly approaches $\gamma t$, but, at the same time,
\mbox{${dR \over dt}|_{t=0}=0$,} and
initially, the second time derivative of $R(t)$ follows 
$-\langle \dot{X}(t) \dot{X}(0) \rangle$.
Such an ansatz is given by the
formula:
\begin{equation}
R(t) =  
\hbox{exp}\left[ - \int_0^t (t - t^{\prime}) 
\langle \dot{X}(t^{\prime}) \dot{X}(0) \rangle dt^{\prime} \right].
\label{exp_ansatz}
\end{equation}
In the context of the exchange narrowing problem, 
this formula  was first obtained by Anderson and Weiss\cite{AW}
on the basis of a Gaussian random noise model.

Fast decaying correlation functions like 
$\langle \dot{X}(t) \dot{X}(0) \rangle$ are usually difficult to compute, 
but they can be
either  reasonably approximated or, sometimes, extracted from 
experiment. Extensive experience with the Green-Kubo formula leaves
no doubt that it is a reliable quantitative recipe provided the
long-time Markovian assumptions are adequate.  It is, however, clearly
seen from the above discussion that, if the assumption of smooth
approach to the Markovian behavior is incorrect, then the Green-Kubo
formula should be abandoned.

\subsubsection{General argument}
\label{gen_arg}

Now we present a general intuitive argument that indeed, as the above
discussion suggests, there should be no  persistent short-time
fluctuations  of the correlation function $R(t)$ around the exponential
Markovian behavior.

Let us return to Assumption~1 made in Section~\ref{exponential},
which
postulated that $R(t)$  can be treated as a relaxation function.
A subtlety associated with that assumption is that, 
on the one hand, it is given in terms of the quantity 
$\langle X(t) \rangle_{\rho_{_R}(0)}$ requiring the averaging
over the ensemble of the sample evolutions of the entire system,  
but, on the other
hand, 
the same quantity is equally expected to represent just one sample
evolution of the
macroscopic system in the sense of a real experiment, in which
the initial conditions are randomly set in compliance with the
initial probability distribution $\rho_{_R}(0)$, 
and then the value of $X(t)$ is measured.
This can only be true, if $X(t)$ is 
--- or can
be considered as --- either
the sum or   the average of infinitely many equivalent 
contributions from statistically independent parts of the system.
Such a condition is, normally, fulfilled for any quantity accessible
by a macroscopic measurement.

In the following, we assume that the variable $X$ is
decomposable as stated above, i.e. 
\begin{equation}
X(t) = \sum_m \tilde{x}_m(t),
\label{decompose}
\end{equation}
where $\tilde{x}_m(t)$ denotes one of many equivalent contributions
to $X(t)$.
Therefore, the correlation function $R(t)$ can be rewritten as
\begin{equation}
R(t) \simeq \langle \tilde{x}_m(t) \rangle_{\rho_{_R}(0), m},
\label{R1}
\end{equation}
where the additional average is taken over all trajectories 
$\tilde{x}_m(t)$ during the same sample evolution of the entire system. 

In Section~\ref{exponential},  
the variable $X$ was assumed
to be the only slow variable characterizing the large system --- therefore,
the dynamic evolution of the variables $\tilde{x}_m$ is
supposed to exhibit fast fluctuations  representative of the short
(ballistic) timescale of the problem.

\begin{figure}
\setlength{\unitlength}{0.1cm}
\begin{center}
%=======================================================================
\begin{picture}(82, 52)
{
\put(-10,-13){

\epsfxsize=4.5in
\epsfbox{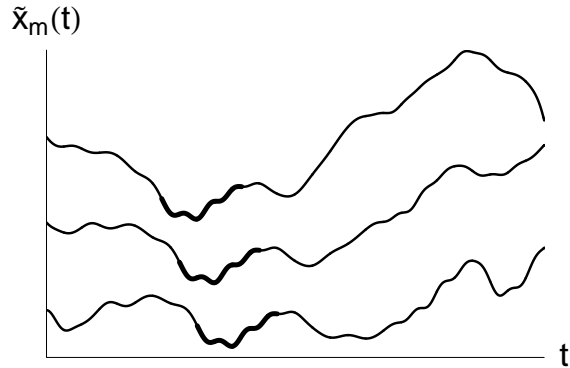} }
}

\end{picture} 
%============== 
\end{center}
\caption[\ Identical short time pieces of different trajectories]
{This picture illustrates the argument made in the text that
identical short-time pieces can appear in    
different trajectories
of variable $\tilde{x}_m$. 
The three trajectories  shown in this picture have
one identical  piece $-$ the thicker part of each trajectory line.
These trajectories have been displaced along the  vertical axis 
to make them distinguishable. 
In this picture, the identical pieces are 
shifted in time by an interval which is  shorter than the time length
of each piece. In general, the time shift between
the identical pieces can be both
very short and very long.

} 
\label{trajectories} 
\end{figure}

Considering  the  ensemble  averaged  quantity     
$\langle \tilde{x}_m(t) \rangle_{\rho_{_R}(0), m}$,
it is important to realize
that any violent short-time event experienced by
one trajectory $\tilde{x}_m(t)$ can be experienced simultaneously   by
many other trajectories.  
Moreover, another set of trajectories
$\tilde{x}_m(t)$ can also 
go through an identical short-time event,
shifted in time by an interval much shorter or much longer 
than the scale of this
short-time event itself (see Fig.~\ref{trajectories} for the
illustration). In other words,  any given short-time event
can start at any given moment of time.
Therefore,  in order for the short timescale to be pronounced after
averaging over all trajectories, it is necessary that not only the
short-time events were present in each trajectory, but also the
probability of those events must fluctuate on the same short
timescale. 

The short-time fluctuations around the Markovian behavior of ensemble averaged
quantities are absent, because
the probability of a given short-time event is determined
by a very large number of random contributions from uncorrelated
trajectories. Being an extremely well averaged quantity, the above
probability does not fluctuate.
Even if 
the  fluctuation of the probability of a specific short-time event 
accidentally occurs, then there are
many different short-time events to be averaged over, and it is very
unlikely, that the probability fluctuations for all of them happen
at the same time and affect the average in the same direction.

To make the above discussion complete, one would  have to answer
the question: What is the difference between the initial
probability distribution $\rho_{_R}(0)$ and the probability distributions at
later times  that explains why the short-time deviation of $R(t)$ from
the slow Markovian behavior occurs only around $t=0$?  This question  
will be addressed in
Sections~\ref{hyperbolic} and \ref{conj_limits} in a more
general context.

\subsubsection{Conjecture I}
\label{conjecture}

The preceding discussion motivated the postulate that whenever a kind
of  Markovian description is assumed to be adequate for long space- and
timescales, the functional form of the correlation functions obtained
in the framework of such a description also applies to the short timescale,
with the only exception being an initial time interval of the order of
the mean free time.
As far as the corresponding relaxation process is
concerned, then it is difficult to 
embrace that postulate without concluding that, after a
certain time from the beginning of the nonequilibrium evolution,
not only the resulting Markovian functional form but the Markovian rate
equations themselves apply to the short timescales, and, if necessary,
the spatial coarse-graining for the Markovian description can also be
chosen very fine.

Making one step from the above seemingly formal observation, we
now introduce a stronger statement:

{\it Conjecture I}: 
We consider a many-body system relaxing to equilibrium  
in the linear response regime and conjecture that in this process,  
after  the memory of the initial probability
distribution is lost,  a kind of randomness subsequently  
propagates to  very short space and time scales.  This randomness is
such that, considering the nonequilibrium behavior of the {\it ensemble
averaged} quantities, it is appropriate to
invoke a Brownian-like Markovian description,
which is  consistent with the long-time Markovian assumptions,
but, at the same time, based on a very fine coarse-graining 
with the scale  {\it much smaller} than the scale  of the
ballistic behavior of the particles composing the large system. 
(The
precise meaning of the  
``Brownian-like Markovian description'' will be
clarified in Section~\ref{cl-cont} in a system-specific context.)

Two closely related aspects make the above conjecture 
different from the previous treatment:
First, it does not require 
the quantity of interest to evolve much slower than a typical
microscopic variable.
Second, it does not call for
the long-time Markovian description to be developed first
and then extended to the short timescales --- it only requires
the long-time Markovian assumptions to be consistent with
the resulting description.

The last circumstance makes {\it Conjecture I} complementary to the
long-time assumption of chaotic mixing for the classical spin
trajectories on a sphere (Section~\ref{cl-intro}).  As we have mentioned in
Section~\ref{cl-intro}, it is impossible to  construct a self-contained
long-time description on the spherical surface because of the physical
absence of distances much greater than the mean free path.  (The
mean free path in our spin problem is of the order of the radius of the
sphere).  However, if the spherical surface is coarse-grained in very fine
elements, then, locally, those elements can be considered flat. It is,
therefore, natural to assume, that the random properties on the very
fine scale in the case of random motion on the spherical surface are
not qualitatively different  from the case of Brownian motion on the
infinite plane. Therefore, the rate equations similar to those
used in Section~\ref{1d-brownian} can be introduced on the spherical
surface.  An analogous situation will characterize the quantum case in
Section~\ref{quantum}.

\subsubsection{Local phase space picture for hyperbolic-like systems}
\label{hyperbolic}

In this part, we introduce a complementary line of arguments
involving the hypothesis that the many-body
system addressed by {\it Conjecture I} is a chaotic hyperbolic system 
\cite{Gaspard}. The connection between {\it Conjecture I}
and the hypothesis of hyperbolicity is not an obvious one.
We believe that hyperbolicity is not a necessary
condition for {\it Conjecture I}, and we do not even
speculate whether the hyperbolicity is sufficient 
to prove {\it Conjecture I}. 
In fact, only a very minimal consequence of  hyperbolicity,
the expansion-contraction picture, will be used in
our arguments, and only  in conjugation with ``physical''
(albeit mathematically uncontrollable) assumptions.
Given all the above reservations, the origin and the limits of
{\it Conjecture I} should, nevertheless, 
become more transparent when discussed in the context
of hyperbolic systems.

The following discussion deals with
an abstract many-body Hamiltonian system, about which
it is no longer assumed that it
exhibits separation between slow and 
fast motions. For consistency of language, 
the Hamiltonian system in question is assumed to be 
very large but finite. As a prototype of such a system,
we keep in mind the lattice of interacting classical
spins introduced in Section~\ref{formulation}.

In the many-body  phase space of a Hamiltonian system, the phase volume
is conserved under  dynamical evolution. When the whole system shows
chaotic mixing,  it means that any continuous set of points initially
occupying a small part of the phase space would eventually be
distributed over the whole phase space --- subject to the energy conservation
constraint.
(This property is consistent with the
mixing we assumed in Section~\ref{cl-intro} 
for the individual particle trajectories but
somewhat more restrictive.)  A volume-preserving
continuous  procedure of spreading  the above set of points is to
expand a part of this set in certain (unstable) directions and contract
it in other (stable) directions, keeping the volume unchanged.
At a given point of the phase space, 
the stable and unstable directions can be obtained by the
linearization of the equations of motion in the vicinity of that point.
Assuming that the dynamical evolution ensures such an
expansion-contraction (hyperbolic) property for most points and most
directions in the phase space, the initially compact small volume will
be spread over the phase space  as a continuous random pattern of very
thin cells.  The longer is the time  allowed for this process, the smaller
is  the
ultimate transverse size of the thin cells.  This
transverse size corresponds to  the contraction directions of the phase
space. It characterizes the scale of dynamically developed randomness
and thus underlies the propagation of the Markovian assumptions to very
short scales.

At the same time, if a smooth not-too-fast varying probability
distribution is assigned within some initial small volume, such a
distribution will be stretched over much larger phase space, and,
therefore it will become nearly flat along the longitudinal (expansion)
directions of the above cells.  This extreme smoothness of the
random pattern along the expansion directions makes the integrated
average over large subvolumes behave in the same way as the average
over much smaller subvolumes, and, therefore, allows the Brownian-like
description for the long space and time intervals 
to be extended to much shorter
intervals.

\

\begin{figure}
\setlength{\unitlength}{0.1cm}
%=======================================================================
\begin{picture}(100, 40)
{
\put( -2, -127){

\epsfxsize= 4.65in
\epsfbox{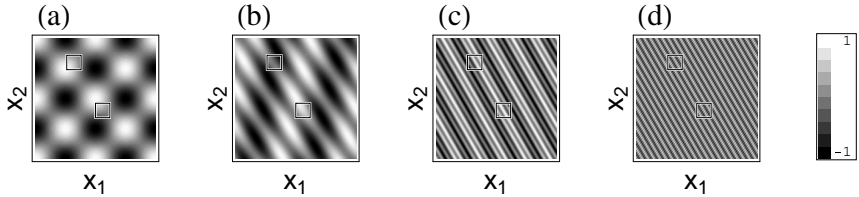} }
}

\end{picture} 
%============== 
\caption[\ Expansion-contraction picture.]
{Expansion-contraction picture \protect{---} an oversimplified illustration
to the arguments given in Section~\protect{\ref{hyperbolic}}.
Coordinates $X_1$ and $X_2$ represent a {\it full} many-body
phase space. The interval shown along each axis is
$(- 2 \pi, 2 \pi)$.  The coloring scheme 
(displayed in the rightmost frame)
encodes function $f(X_1, X_2)$ proportional to
a {\it weak deviation} from an
equilibrium probability distribution. In figure (a),
$f(X_1, X_2) = \hbox{cos} X_1 \hbox{cos} X_2$  \protect{---}
it is an example of a factorized distribution.
In figures (b-d), the distribution function from figure (a)
is expanded by factor $k$  along the direction 
$(- \hbox{cos} (\pi / 3), \hbox{sin} (\pi / 3) )$, and
contracted by the same factor along the perpendicular direction
$( \hbox{sin} (\pi / 3), \hbox{cos} (\pi / 3) )$.
The values of $k$ are the following: (b) $k=2$; (c) $  k=8$; (d) $k=32$.
Two small squares within each of the plots represent two sample coarse-grained
elements discussed in the text.
} 
\label{fig_expand} 
\end{figure}

The expansion-contraction picture is visualized  in Fig.~\ref{fig_expand}\cite{ZBD}.
Although the probability distributions presented  in that figure 
do not exhibit chaotic patterns, they are, nevertheless, sufficient to
illustrate the qualitative difference between the factorized distribution
(Fig.~\ref{fig_expand}(a)) and the distributions emerging at later stages of
the expansion-contraction process (Fig.~\ref{fig_expand}(d)). 
That illustration, however,
requires the following explanation.

Let us assume that a fine coarse-graining is introduced in
Fig.~\ref{fig_expand} in the space of variables ($X_1$, $X_2$), and then
consider two coarse-grained elements indicated in that figure.  These
two elements are  displaced from each other along the expansion direction.  
Initially (Fig.~\ref{fig_expand}(a)),
the averaging over each of them
would give two probabilities substantially different from each other.
At later stages of the expansion-contraction process 
(Fig.~\ref{fig_expand}(d)), the two coarse-grained
elements become connected via the cells of extremely smooth probability
distribution, and, therefore, the averaging over 
these elements would produce
nearly identical results. 

It might appear from Fig.~\ref{fig_expand} that the generic situation
is the one, where two coarse-grained elements are displaced
not along the expansion direction but along an arbitrary one.
In that case, the two elements will not be connected
via cells of smooth distribution. 

The above objection, however,
is based on an artifact of two-dimensional illustrations.
A detail, which is difficult to represent graphically,
is that the phase space of a system with a very large number of particles
is many-dimensional, while, normally, the coarse-graining is applied
to a few-dimensional subspace of that space 
(e.g. the subspace of one-particle coordinates). 
When a many-dimensional probability distribution
exhibits an expansion-contraction 
pattern, the crucial difference of that pattern 
from the one presented in Fig.~\ref{fig_expand} is that
the number of expansion directions is supposed to be much greater
then the dimensionality of the coarse-grained subspace of interest. 
(In Fig.~\ref{fig_expand}, the number of expansion directions is one, and
the dimensionality of the coarse-grained space is two.)
If a  many-dimensional
cell of smooth distribution has a projection on a given
a few-dimensional coarse-grained element, then such
a projection is overwhelmingly likely to be smoothly extended
to all coarse-grained elements surrounding the given one.

\

We can now explain why {\it Conjecture I} distinguishes  the initial
and the long-time behavior of the correlation functions.

From the viewpoint of the expansion-contraction picture, 
the small parameter implicitly underlying {\it Conjecture I} is
the ratio of the characteristic scales of the many-body probability
distribution along the contraction  and expansion directions.  
Such a small parameter develops dynamically
--- it is not present in the initial probability distributions of a typical
nonequilibrium problem. 

As an example, let us consider our nonequilibrium 
problem for classical spins,
in which case the initial probability
distribution $\rho(0)$ is given by Eq.(\ref{rho}).
According to Eq.(\ref{rho}), the probability of initial polarization of
each spin is completely independent of the polarizations of its
neighbors, which means that $\rho(0)$  can be factorized in terms of
the distributions describing each spin separately.  In essence,
this factorization
implies that a single scale (that of the initial one-spin
distribution) characterizes the many-spin distribution along all
directions of the phase space. That scale corresponds to the radius of
the sphere on which the one-spin distribution is defined. (The explicit
form of this distribution will be given in Section~\ref{cl-cont}.) Thus, as
always happens with well-defined initial conditions of many-body
problems, the initial distribution (\ref{rho}) is not complex enough to
discriminate between the contraction and expansion directions intrinsic
for the many-body phase space of a given dynamical problem.  Therefore,
some time ($\sim \tau$) is required before the chaotic pattern
possessing the necessary small parameter is established
(see Fig.~\ref{fig_expand}).

\subsubsection{General scope of Conjecture I}
\label{conj_limits}

Returning to the argument given in Section~\ref{gen_arg}, 
it is
important in general, --- and for the treatment of the quantum case in
particular --- that the argument is not limited to the case when the
trajectories in question describe a subsystem within a much larger chaotic
Hamiltonian system.  In general, this argument applies when  a subsystem is
subjected to the influence of an external environment which follows
continuous dynamics and has a continuous distribution.  As long as 
the long-time mixing assumption is satisfied, 
the argument does not
exclude a non-chaotic or non-Hamiltonian environment, and neither
does {\it Conjecture I}.

In the case of a non-chaotic or non-Hamiltonian environment, 
the extreme stretching of
the subsystem-plus-environment probability distribution 
(similar to the one described in Section~\ref{hyperbolic})
should also underly {\it Conjecture I},
though, in such a case, it is more difficult to visualize the small parameter.
Nevertheless, it is clear that {\it
Conjecture I} $\;$ becomes applicable after the 
 probability distribution
develops the complexity intrinsic to a given dynamics.

\subsubsection{One-dimensional Brownian motion}
\label{1d-brownian}

In order to facilitate the Brownian-like description for classical spins in 
Section~\ref{cl-cont}, 
here we describe the standard
one-dimensional diffusion of Brownian particles in a fashion similar
to the treatment of the exponential decay in Section~\ref{exponential}.
Instead of one slow variable considered in Section~\ref{exponential}, the
treatment of  diffusion requires a continuum of slow variables.

In the problem of Brownian motion\cite{Einstein}\cite{Kubo}, 
the Markovian timescale should be
associated with a slow change in the position coordinate $x(t)$ 
of a Brownian
particle. The drift of Brownian particles is, usually, characterized by
the correlation function 
\begin{equation}
B(t) = \langle x^2 \rangle - \langle x(t)x(0) \rangle .
\label{B_def}
\end{equation}
We consider the problem as if many Brownian particles move
simultaneously, and the average is taken over all of them.

In order to proceed with the Markovian description,
it is necessary to introduce   spatial
coarse-graining with
the size $\Delta_x$  
of the coarse-grained subvolumes  (linear intervals
in this case). This size is   much greater than the
mean free path of the Brownian particles.
The slowly changing  particle populations
in  each subvolume   
become the primary variables of the
Markovian description. 
It is then assumed that
the Markovian equations should only describe the direct exchange
of particles between adjacent subvolumes, which gives
\begin{equation}
{ \partial f(t, x_{(i)}) \over \partial t} 
= W  \sum_j (f(t, x_{(j)}) - f(t, x_{(i)})), 
\label{rate_eq}
\end{equation}
where $f(t, x_{(i)})$ is the nonequilibrium fraction of the particle
population in subvolume number $i$; the sum over $j$ includes only the
subvolumes adjacent to the $i$th subvolume;
and  $W$ is the rate of transitions across the boundary
of two adjacent subvolumes.
There are two additional assumptions behind Eq.(\ref{rate_eq}). Namely:
(a) every
Brownian particle is always affected by the same equilibrated
environment; and (b) the transitions in both directions ($(i) \to (j)$ 
and $(j) \to (i)$) are equivalent.
The continuum limit of the description represented by 
Eq.(\ref{rate_eq}) 
yields the standard
diffusion equation:
\begin{equation}
{ \partial f(t, x) \over \partial t} 
= D \; { \partial^2  f(t, x) \over \partial x^2} , 
\label{simp_dif}
\end{equation}
where $D$ is the diffusion coefficient equal to $ W \Delta_x^2 $.

The solution of Eq.(\ref{simp_dif}) allows $B(t)$ to be calculated,
which gives
\begin{equation}
B(t) =  D t. 
\label{B}
\end{equation}

According to the general recipe preceding Eq.(\ref{G-K}),
the Green-Kubo expression for the diffusion coefficient can be obtained by
equating
the first time derivative of the Markovian result for
$B(t)$  to
the saturation value of the integral of 
the exact second time derivative of  $B(t)$. 
After a manipulation, which is similar to the one
as a footnote in Section~\ref{exponential},
the second derivative of $B(t)$ can be presented as 
$\left\langle v(t) v(0) \right\rangle$,
where $v(t) = {dx(t) \over dt}$. 
Therefore, the expression for the diffusion coefficient becomes
\begin{equation}
D = \int_0^{\infty} \left\langle v(t) v(0) \right\rangle dt.
\label{D}
\end{equation}

Thus we have again illustrated the general statement that the
assumption of the smooth approach to the Markovian behavior is
sufficient in order to
obtain the standard Green-Kubo results for the parameters describing
the Markovian approximation.

\subsection{Classical spins --- continued}
\label{cl-cont}

{\it Conjecture I} can now be applied to the system of classical spins 
in the following way.

Let us assume that, on the spherical surface corresponding to the $n$th
spin, we trace a large number of sample trajectories of that spin.
Different sample trajectories of the $n$th spin correspond to different
realizations of the dynamical evolution of the entire system, and the
initial conditions for each such a realization are chosen in accordance
with the initial probability distribution (\ref{rho}).  We thus have a
large number of sample points moving simultaneously on the spherical
surface.

Now let us coarse-grain the spherical surface in sufficiently small elements. 
The greater is the target
accuracy of the calculation, the finer is the coarse-graining.
The large number of 
sample trajectories should guarantee that, at any moment of time, there
are enough sample points in each coarse-grained subvolume (surface
element, in this case) to give an accurate representation of the
probability to find a point in this subvolume.

The probability distribution of the sample points
on the $n$th sphere is initially close and expected to remain close to
the uniform distribution corresponding to the infinite temperature
equilibrium. 
We denote the small deviation from the uniform 
distribution as $f_n(t, x_n)$, 
where $x_n$ stands for two standard  spherical angles $\theta_n$
and $\varphi_n$, chosen such that
$S_n^\mu = S \; \hbox{cos} \theta_n$, where $\mu$
is the same as the one entering Eq.(\ref{G}). 
The expansion of $\rho(0)$
in powers of $\beta_0$ gives
\begin{equation}
f_n(0, x_n) \simeq \beta_0 \; 
\hbox{cos({\boldmath $q \cdot r$}$_n$)}
\;  S \; \hbox{cos}  \theta_n .  
\label{initial}
\end{equation}

We now recall, that,
in Section~\ref{cl-intro},  we have already introduced 
the notion of the mean free path
of a typical sample point on the above spherical surface. 

When the coarse-grained scale is smaller than that mean free path,
the influx and outflux events for the subvolumes
along a given trajectory  will be correlated in time.  However, for
a given subvolume, there should be many sample points entering and
leaving it within any finite time interval. In equilibrium, 
the non-Markovian
influx and outflux events should be perfectly balanced by the exact
dynamics of the system.  It is the slight imbalance between the influx and
the outflux events that governs the nonequilibrium evolution in that
subvolume. 
The property  underlying {\it Conjecture I} can now be
reformulated as follows:  
At the later stages of the ensemble's evolution, there are no important
correlations between the 
uncompensated influx and outflux events for a given coarse-grained
subvolume at the given and at the previous moments of time. The time 
interval between the two moments in question 
can be substantially shorter than the time
required for a sample point to cross that subvolume.

In the long-time regime describable by {\it Conjecture I}, 
the 
population of sample points in a given subvolume becomes a
``slow'' variable. In {\it Conjecture I},
by postulating the applicability of Brownian-like description,
we implied that the problem can now be treated as if
the population of the sample points in each subvolume
describes Brownian particles
with mean free path much smaller than the size of that subvolume.

If the $n$th 
spin was in an equilibrium environment, then the
description of Brownian motion presented in Section~\ref{1d-brownian}
could be adopted, and  the continuum limit of such a
description would lead to a diffusion equation similar to
Eq.(\ref{simp_dif}).  However, in
our case, the environment, which means spin distributions on the
neighboring lattice sites, is slightly out of equilibrium.

At the level of rate equation (e.g. Eq.(\ref{rate_eq})),
the leading order effect caused by
the nonequilibrium environment is
that the transition rate from one subvolume to another becomes slightly
different from the transition rate in the opposite direction.  In that
case, the {\it equilibrium} fractions of particles in the two
subvolumes respond to the disbalance of the transition
rates and create the additional particle flux between those
subvolumes. 
Below, we proceed directly to the continuum description, which takes into
account the above effect.

In order to avoid unnecessary details, we adopt a schematic
notation, which makes all quantities look one-dimensional,
but implies the proper number of dimensions and the proper covariant
form of the differential operations.  In such a notation, the 
linear response formula for the probability flux  on the $n$th site can
be written as 
\begin{eqnarray} 
j_n(t, x_n) \;  & = & \;  - D_n(x_n)
{\partial f_n(t, x_n) \over \partial x_n} 
\label{flux} \\ \nonumber
 & & + \;
\sum_{k} \int_{V_{x_k}} K_{nk}(x_n, x_k) f_k(t, x_k) dx_k,
\end{eqnarray} 
where $D_n(x_n)$ is the diffusion coefficient (actually, tensor)
corresponding to the standard diffusion in the equilibrium
environment; 
$K_{nk}(x_n, x_k)$ is the kernel of the term representing the linear flux
response  on the $n$th site to the slight deviation from the equilibrium
distribution on the $k$th site (this response is analogous to Ohm's law); 
and $V_{x_k}$ stands for the entire space
of variables characterizing the $k$th site.

The probability density $f_n$ and the probability flux $j_n$ must
also satisfy the
continuity equation 
\begin{equation} 
{\partial f_n(t, x_n) \over \partial t} = - \hbox{div}[j_n(t, x_n)].  
\label{dfn} 
\end{equation}
In general, pairs of equations (\ref{flux}, \ref{dfn}) should be
written for each lattice site, and together they form a closed set of
integro-differential equations. 

Although introduced above as a natural development of the previous
treatment, the integral term in Eq.(\ref{flux}) 
represents quite a dramatic step in the
extension of the Brownian motion formalism beyond the limits of the
conventional Markovian approximation. This term should absorb the seemingly
intractable dynamical correlations between different spins.
A simple picture behind that term is the following:
The nonequilibrium probability of a certain orientation of the $k$th
spin creates a preferred direction of the local field by which the
$k$th spin affects the $n$th spin. In turn, 
the preferred direction of the local
field means the preferred direction for the additional
probability flux on the $n$th spherical surface. 

In order to promote the Brownian-like
description consistently, 
it is necessary to assume that the response of one slow
variable (the population of a given subvolume)
to the deviation  of another slow variable from the equilibrium can 
be caused not only by the direct interaction between 
those variables but also by the indirect effects mediated by other
spins. This
means that there can exist a nonzero  kernel $K_{nk}(x_n, x_k)$
coupling the $n$th and the $k$th spins, even though those spins do not
interact directly. Another kernel not to be neglected 
is $K_{nn}(x_n, x_n^{\prime})$. 
In general, it is only reasonable to expect that,
when the distance between the spins becomes much greater than the radius
of interaction, the indirect effects
become insignificant.

The evaluation of $D_n(x_n)$ and $K_{nk}(x_n, x_k)$ from the knowledge
of the Hamiltonian (\ref{H}) is not attempted in this work, and, in
fact, we are not sure if such a task is achievable --- the absence of a 
self-contained long-scale description on a sphere precludes us from
applying a straightforward generalization of
the Green-Kubo recipe (\ref{D}). Nevertheless, as we
show below, it is possible to come to very definite conclusions about
the time-dependences of the resulting solutions.

When the combined symmetry of the Hamiltonian and the initial
conditions is such that 
two initial distributions $f_n(0, x_n)$ and $f_k(0, x_k)$ can be
transformed into 
each other, then the dynamic
evolution induced by the Hamiltonian retains that
equivalence. Thus those two sites can be characterized by the same
probability distribution also in the long-time diffusive regime.

The nonequilibrium problem corresponding to 
the calculation of the correlation function  (\ref{G})
can now be greatly simplified, because in this problem 
all probability distributions $f_n(t, x_n)$  
are equivalent. This equivalence is related
to the fact that the irreducible representations of the
group of lattice translations have form 
$\hbox{exp($i${\boldmath $q \cdot r$}$_n$)}$ (cf. Eq.~(\ref{initial})).

The symmetry argument can be illustrated using the
examples presented in Fig.~\ref{examples}(b). In that figure,
all spins  are explicitly equivalent in example (I).
It is also quite obvious that the probability distributions
corresponding to the alternating spin sites in example (II) are also
equivalent. In the other two examples, one would
have to imagine that the probability distribution of each spin
is the real part of 
a complex-valued function 
which is the same for all spins up to the helically arranged phase.

It thus follows that, in our problem, 
\begin{equation}
f_n(t, x_n)|_{x_n = x_0} = 
\hbox{cos({\boldmath $q \cdot r$}$_n$)} f_0(t, x_0)
\label{equivalent}
\end{equation}
Given the above relationship one can easily obtain from
Eqs.(\ref{flux}, \ref{dfn}) that
\begin{equation} 
{\partial f(t, x) \over \partial t}  =  
{\partial \over \partial x} \left( D(x) {\partial f(t, x) \over \partial x} \; -
\int_{V_{x^{\prime}}} K(x, x^{\prime}) f(t, x^{\prime}) dx^{\prime}
\right) ,
\label{diffusion} 
\end{equation}
where $f(t, x) \equiv f_0(t, x_0)$, and
\begin{equation}
K(x, x^{\prime}) = 
\sum_k \hbox{cos({\boldmath $q \cdot r$}$_k$)} K_{0k}(x,x^{\prime}).
\label{K}
\end{equation}

Equation (\ref{diffusion}) represents correlated diffusion on a
spherical surface. Like an ordinary diffusion equation, this equation has a
set of solutions of the form 
$f_{(\lambda)}(t, x) = e^{- \lambda t} u_{\lambda}(x)$, where $\lambda$  is
one of the eigenvalues  of the integro-differential linear
operator acting on $f(t, x)$ in the right-hand side of 
Eq.(\ref{diffusion}), and $u_{\lambda}(x)$ is the corresponding eigenfunction.

Unlike a typical diffusion problem, the eigenvalues of our problem can
have both real and imaginary parts. This property originates from the
fact that, in general, the kernel $K(x, x^{\prime})$ has no
symmetry with respect to interchange of the variables $x$ and
$x^{\prime}$, and, as a result, the integro-differential operator 
in Eq.(\ref{diffusion}) is non-Hermitian.  

The asymmetry of the kernel
$K(x, x^{\prime})$ can be traced back to the asymmetry of
the  kernels $K_{0k}(x,x^{\prime})$ entering Eq.(\ref{K}). 
Although it is impossible to separate various factors
influencing the kernels $K_{0k}(x,x^{\prime})$, 
it is, at least, clear that those kernels are
strongly affected by the direct spin-spin interaction (\ref{H}). That
interaction is asymmetric in the sense that, in general, the $k$th spin
with orientation $x^{\prime}$ 
creates on the zeroth site a local field, which is
different from the local field created on the $k$th site by the zeroth
spin with some arbitrary orientation $x$.

Since we deal with a finite volume (meaning spherical surface), the
eigenfunctions $u_{\lambda}(x)$ necessarily 
form a discrete set and generate a
corresponding discrete set of eigenvalues.
(If the eigenvalues of our problem are complex, then
only the real part of the solution should be taken.) 
We further assume
that the equilibrium is stable, and, therefore, 
the underlying dynamics of the spin system guarantees that the
real parts of all eigenvalues are non-negative.

In order to obtain $G(t)$ from the above description, one has to
average the $\mu$th polarization of the zeroth spin over 
the solution of
Eq.(\ref{diffusion}). Therefore, 
\begin{eqnarray}
G(t) \; &\simeq& \;  \int_{\hbox{\small sphere}} 
\hbox{cos}\theta  \; f(t, \theta, \varphi) \; 
\hbox{sin}\theta \;  d \theta \; d \varphi 
\nonumber 
\\
&=&
\int_{\hbox{\small sphere}} 
\hbox{cos}\theta  \; \sum_{\lambda} 
e^{- \lambda t} u_{\lambda}(\theta, \varphi) \; 
\hbox{sin}\theta \;  d \theta \; d \varphi
\label{average}
\end{eqnarray}
where we returned to the  spherical variables 
$\{ \theta, \varphi \}$ introduced earlier
(with index $n$ dropped).
As a result, the generic long-time
behavior of $G(t)$  is given by Eqs.  (\ref{longexp},\ref{longcos}) --
it is  controlled by the eigenvalue
that has the smallest real part among those, whose respective
eigenfunctions $u_{\lambda}(\theta, \varphi)$ 
give nonzero contribution to the integral in Eq.(\ref{average}).

The alternatives to the long-time behavior (\ref{longexp},
\ref{longcos}), which we left out as ``non-generic'', correspond to the
following possibilities: (i) the system does not experience chaotic
mixing sufficient to justify the diffusion description (\ref{flux},
\ref{dfn}); (ii)  within the diffusion description, the expansion of
$G(t)$ contains more than one eigenvalue with the
smallest real part (complex conjugates do not count).

\

In a typical case, the problem has only one
characteristic timescale $\tau$ given by Eq.(\ref{tau}).
Therefore, we expect the characteristic
scales of both $D(x)$ and $K(x,x^\prime)$ 
to be given by the appropriate combination of
$\tau$ and the radius of the sphere.  
With this, we can make the following simple estimate of the time, after
which the exponential dependence (\ref{longexp}) or (\ref{longcos})
should become pronounced.

It presumably takes a time of the order of $\tau$ 
to reach the Markovian regime describable by
Eq.(\ref{diffusion}).  In the Markovian regime, both the slowest
exponent $\xi$ entering Eqs.(\ref{longexp}, \ref{longcos}) , and the
difference between $\xi$ and the second slowest exponent should
be of the order of $1/\tau$, implying that it
will take another time of the order of  $\tau$ before the
contributions from the faster exponents become suppressed.
Therefore,  the behavior of $G(t)$ should approach  asymptotic form 
(\ref{longexp}, \ref{longcos}) after a time of the order of several 
$\tau$, i.e. sufficiently fast.

In the context of experimental or numerical verification of the
long-time behavior (\ref{longexp}, \ref{longcos}), the estimate just
given should be complemented by reasonably good luck. For example, if
the difference between the slowest exponent and the second slowest
exponent equals one third of the slowest exponent, then the two
exponents still compete over quite an extended time interval, and by the
time the faster of those exponents becomes completely suppressed, the
overall value of the correlation function becomes too small to be
obtained from experiments or from numerical calculations.

\subsection{Quantum spins}
\label{quantum}

Generalization of the previous treatment to quantum spins requires one
modification, namely, instead of  trajectories of the
tip of the classical spin on a sphere, we consider trajectories in
the space of parameters describing the density matrix of a
quantum spin. 

As straightforward as the above approach might seem, it is not how 
Markovian assumptions are usually introduced in quantum problems.
Usually, Markovian assumptions are based either on the classical
limit of the quantum problem, or on classical models for the
occupation numbers of the quantum states (e.g. Ising model). Both
of those approaches are deficient because of the lack of invariance
with respect to the transformations of the Hilbert space of the quantum
problem. Sometimes (not in our case), such a deficiency appears
to be unimportant, but
it always makes the discussion of compatibility of 
the Markovian assumptions with the exact dynamics 
intractable even at the level of posing the problem.
If the Markovian assumptions are to be discussed in a situation where
their extreme implications are expected to be quantitatively adequate,
it is crucially important to have those assumptions grounded on the
well-defined notion of trajectories consistent with the symmetry of
the Hilbert space.

A more recent trend is to relate the notion of  chaos
in quantum systems 
to the Wigner-Dyson  statistics of energy levels.
(See e.g. Refs.~\cite{Poilblanc,PG}.)
In this work, however, the Wigner-Dyson statistics
is neither assumed nor derived from another assumption.
We can
only remark that the spin~1/2 $XXZ$ 
chains discussed in
Section~\ref{empirical} do not exhibit
the Wigner-Dyson statistics\cite{Poilblanc}
but show the long-time behavior (\ref{longcos}).

Below, in order to be specific, we limit our treatment 
only to spins 1/2. We also assume that the
spin system is very  large but finite, and, therefore, the
number of states in the quantum basis is also finite.

The $2 \times 2$ density matrix of a given spin 1/2 can always be
obtained by averaging over the full many-body density matrix.  
Already an ensemble average, the one-spin density matrix is expected to
show, not a random behavior, but, in our case, a rapid relaxation toward
the $2 \times 2$ unit matrix corresponding to the infinite temperature
equilibrium.  Therefore, in order to identify the underlying chaotic
trajectories, we have to trace back the quantum mechanical and statistical
averaging. 

We propose a description that treats different spins by using different
basis sets of the many-body wave functions.  Considering the $n$th spin,
we adopt  an interaction-like
representation for the basis wave functions:  First, at $t=0$, for the whole 
system excluding the
$n$th spin, we choose a complete orthogonal set of  wave functions $
\{ \Psi_{n\alpha}(0) \}$ enumerated by index $\alpha$. Each wave
function $\Psi_{n\alpha}(0)$ is such that every spin has  definite
projection on the axis $\mu$ entering
Eq.(\ref{rho}). (The spins are weakly
polarized along that axis at $t=0$.)
Allowing each of the above wave functions to evolve only under the action
of the part of the Hamiltonian (\ref{H}) not involving the $n$th spin,
we obtain a time-dependent basis set $ \{ \Psi_{n\alpha}(t) \}$. Then,
we choose {\it time-independent} basis wave functions for the $n$th spin as
$\{ \mid \uparrow \rangle , \mid \downarrow \rangle\}$ with the
quantization axis along the same $\mu$-direction of the initial average
polarization of this spin. Finally, the basis set for the whole system
becomes $\{ \; \{ \mid \uparrow \rangle  \Psi_{n\alpha}(t) \} , \;
\{\mid \downarrow \rangle \Psi_{n\alpha}(t) \} \; \}$. 

According to the above recipe, the basis sets designed for the description
of different
spins are identical at \mbox{$t=0$}, but then each of those  sets
evolves differently. The advantages of
this representation will become clear after the formal
structure of our treatment is developed.

In the basis designed for the $n$th spin, the density matrix of that spin 
is just
the average over \mbox{$2 \times 2$} blocks of the density matrix of the
whole system --- each block corresponds to a
fixed value of index $\alpha$.  However, these blocks cannot yet be
treated as
elementary dynamical quantities, because the initial
many-body density matrix (\ref{rho}) is, by itself, 
the result of statistical averaging,
implying  initial thermal contact with the environment.

Since, in our problem, there is no contact with the environment at
later moments of time, we represent the evolution of the
density matrix of the whole system as the average over the evolutions
of many density matrices of ``pure states''.
Pure states are the states of the isolated spin system --- 
each describable by   
a wave function. 
The choice of the ensemble of pure states (to be specified later) 
is not unique, 
because, provided the averaging over the pure state density
matrices at $t=0$ gives the density matrix (\ref{rho}), the result of
the averaging at later moments of time must be independent of other
details of this choice.

The  time-dependent $2 \times 2$ density matrix of the $n$th spin 
can now be considered as an average over $2 \times 2$
time-dependent blocks, each originating from some pure state density
matrix.  If one of
the  pure state wave functions is represented by a set of coefficients
$\{ \; \{  C_{\uparrow n\alpha}(t) \} , \;  \{  C_{\downarrow n\alpha}(t)
\} \}$ in the basis 
\mbox{$\{ \; \{ \mid \uparrow \rangle  \Psi_{n\alpha}(t) \} , \;
\{\mid \downarrow \rangle \Psi_{n\alpha}(t) \} \; \}$}, 
then the elementary block
participating in the averaging is
\begin{equation}
\left(
\begin{array}{cc}
C_{\uparrow n\alpha}(t) C_{\uparrow n\alpha}^\ast (t) \ &
C_{\downarrow n\alpha} (t) C_{\uparrow n\alpha}^\ast(t) \\ [0.2cm]
C_{\uparrow n\alpha}(t) C_{\downarrow n\alpha}^\ast (t) \ &
C_{\downarrow n\alpha}(t) C_{\downarrow n\alpha}^\ast (t) \\
\end{array}
\right).
\label{block}
\end{equation}

Each block of the form (\ref{block}) can be described by
three independent variables: $ \{|C_{\uparrow
n\alpha}|, |C_{\downarrow n\alpha}| ,\Phi_{n\alpha} \}$, 
where
$\Phi_{n\alpha}$ is the difference between the phases of
$C_{\uparrow
n\alpha}$ and $C_{\downarrow n\alpha}$.
Therefore, the time evolution of each of those blocks 
can be mapped to a trajectory of the tip of a Bloch vector
in the three-dimensional space  corresponding to the above set of variables.
The trajectories thus defined become the primary objects in the subsequent
treatment.
We shall call them ``block trajectories''.

In the following,
when no distinction between $C_{\uparrow n\alpha}(t)$ and 
$C_{\downarrow n\alpha}(t)$ has to be made, we shall use
the notation $C_{\uparrow (\downarrow) n\alpha}$. 
We shall also use variables 
\{$|C_{\uparrow n}|$, $|C_{\downarrow n}|$, $\Phi_{n}$\} 
(the same as $ \{|C_{\uparrow n\alpha}|,
|C_{\downarrow n\alpha}| ,\Phi_{n\alpha} \}$ but without index
$\alpha$), whenever we discuss one block trajectory as a representative
of the statistical properties of all trajectories with different
$\alpha$. In particular,  we introduce the probability distribution
$P_n(t, |C_{\uparrow n}|, |C_{\downarrow n}|, \Phi_{n})$, which implies
averaging over the whole ensemble of the pure states and over all
values of the index $\alpha$.  This probability distribution will play
the same role as the probability distribution on a sphere played
earlier in the case of the classical spins.

We can now explain two advantages of the ``interaction representation'' basis 
$\{ \; \{ \mid
\uparrow \rangle  \Psi_{n\alpha}(t) \} , \;  \{\mid \downarrow
\rangle \Psi_{n\alpha}(t) \}  \; \}$.

The first advantage is that, in this representation,  interactions which do
not affect the $n$th spin directly also do not have a direct
influence on the evolution of the block trajectories  ---
those interactions are mostly absorbed by the time dependence of 
$\Psi_{n\alpha}(t)$. The time dependence of the coefficients
$C_{\uparrow (\downarrow) n\alpha}(t)$ is directly controlled
only by those terms in the Hamiltonian (\ref{H}) that include the $n$th spin.  
If the wave functions \mbox{$\{ \; \{ \mid
\uparrow \rangle  \Psi_{n\alpha}(0) \} , \;  \{\mid \downarrow
\rangle \Psi_{n\alpha}(0) \}  \;  \}$} were used as a permanent basis set,
then the
interactions between two spins very distant from the $n$th spin 
would have immediate effect on the coefficients 
$C_{\uparrow (\downarrow) n\alpha}$, which does not affect 
the final
average for the $n$th spin 
but introduces irrelevant fast timescales in the behavior
of those coefficients. 

The second advantage of the interaction-like representation
is that  
each of the wave functions $ \Psi_{n\alpha}(t)$ can be considered
as representing a
generic evolution of the environment of the $n$th spin, and, therefore,
as far as the properties of that spin are concerned, 
it is sensible to
use one probability distribution 
$P_n(t, |C_{\uparrow n}|, |C_{\downarrow n}|, \Phi_{n})$
to describe 
all block trajectories 
corresponding to different values of the index $\alpha$.

Below we specify the initial probability distribution of the coefficients
$C_{\uparrow (\downarrow) n\alpha}$
for the ensemble of many-body pure
states. As we have already mentioned,
this distribution is only constrained by the requirement that
the average of
the initial pure state density matrices
is equal to  the density matrix $\rho(0)$ given
by Eq.(\ref{rho}). We impose one additional constraint 
necessary for the consistency of our treatment: Namely, we require that
the initial probability distribution 
$P_n(0, |C_{\uparrow n}|, |C_{\downarrow n}|, \Phi_{n})$ in the space of
block trajectories  should be such
that it is only slightly different from the ``equilibrium'' probability
distribution $P_n(\infty, |C_{\uparrow n}|, |C_{\downarrow n}|,
\Phi_{n})$. 

Since the object of our interest
is the probability distribution
$P_n(0, |C_{\uparrow n}|, |C_{\downarrow n}|,
\Phi_{n})$ 
rather than the full probability distribution of the pure states,
the details characterizing the latter will be given only to the extent
sufficient to define the former.

The fact that the initial density matrix (\ref{rho}) is diagonal
in the ``initial'' basis $\{ \; \{ \mid
\uparrow \rangle  \Psi_{n\alpha}(0) \} , \;  \{\mid \downarrow
\rangle \Psi_{n\alpha}(0) \}  \;  \}$ allows us to assign equal
probability to all values of the complex phases of the coefficients 
$C_{\uparrow (\downarrow)
n\alpha}$, which implies that the probability distribution
with respect to $\Phi_{n\alpha}$  
is uniform for each value of $\alpha$. Therefore,
\begin{equation}
P_n(0, |C_{\uparrow n}|, |C_{\downarrow n}|, \Phi_{n}) = 
{1 \over 2 \pi} \; p_n(|C_{\uparrow n}|, |C_{\downarrow n}|),
\label{P_n}
\end{equation}
where $p_n(|C_{\uparrow n}|, |C_{\downarrow n}|)$ is the distribution
of $|C_{\uparrow (\downarrow)n}|$ defined by the above equation.

The set of absolute values $\{|C_{\uparrow
(\downarrow)n\alpha}|\}$ for a sample pure state can be considered as a
vector in a high-dimensional space. Since $\sum_{\uparrow
(\downarrow)\alpha} |C_{\uparrow (\downarrow)n\alpha}|^2 = 1$, the tip
of that vector is restricted to a ``quadrant'' of the high-dimensional
hypersphere, where each point has a non-negative projection on each axis.  We
define the probability of sampling on this ``spherical quadrant'' to be
uniform, with small corrections reflecting the weak polarizations of
spins. We do not define those corrections explicitly, because,
as we show below, there is a
simple direct way to see what the resulting distribution 
$p_n(|C_{\uparrow n}|, |C_{\downarrow n}|)$ must be.

Let us first obtain $p_n(|C_{\uparrow n}|, |C_{\downarrow n}|)$ in the case
when there are no small corrections, which corresponds to infinite 
temperature. We denote such a distribution as $p_{n\infty}$. 

According to the above sampling procedure, the probability distribution
$p_{n\infty}(|C_{\uparrow n}|, |C_{\downarrow n}|)$ is
an average over identical probability distributions 
for each of the coefficients  $C_{\uparrow (\downarrow)n\alpha}$. 
The probability for any of the coefficients 
$C_{\uparrow (\downarrow)n\alpha}$ to have a value
in the interval between $y$ (an auxiliary variable) 
and $y + dy$ is proportional
to the ``area'' of the thus-restricted  hypersurface, which is given by
$A_{(N_{\hbox{\footnotesize b}}-1)}(\sqrt{1 - y^2}) dy/\sqrt{1-y^2}$,
where 
$N_{\hbox{\small b}}$ is the number of the states in the quantum basis of the
entire system, and  $A_{N_{\hbox{\footnotesize b}}-1}(\sqrt{1 - y^2})$ is
the surface area of the $(N_{\hbox{\small b}}-1)$-dimensional hypersphere of
radius $\sqrt{1 - y^2}$. Since 
$A_{(N_{\hbox{\footnotesize b}}-1)}(\sqrt{1 - y^2}) \simeq 
(\sqrt{1 - y^2})^{(N_{\hbox{\footnotesize b}}-2)}$,
and $N_{\hbox{\small b}}$ is supposed to be very large,  
the overall dependence of the probability distribution  on $y$ 
should be well
approximated by exp$(- {N_{\hbox{\small b}} \over 2} y^2)$.
As a result,
\begin{equation}
p_{n\infty}(|C_{\uparrow n}|, |C_{\downarrow n}|)  = 
{\pi \over 2 N_{\hbox{\small b}}}
\hbox{exp} 
\left[ - {N_{\hbox{\small b}}  \over 2} 
\left( |C_{\uparrow n}|^2 + |C_{\downarrow n}|^2 \right) \right].
\label{p_inf}
\end{equation}
Predictably, 
the mean value of $\langle
|C_{{\uparrow}(\downarrow)n}|^2 \rangle$
corresponding to $p_{n\infty}(|C_{\uparrow n}|, |C_{\downarrow n}|)$ 
is equal to the inverse number of  basis states.

When the inverse temperature $\beta_0$ appearing in Eq.(\ref{rho}) 
is small  the initial
probability distribution for the $|C_{{\uparrow}n}|$ should correspond
to  infinite temperature for ``spins-up'' and be Gaussian, with  mean
value $\langle |C_{{\uparrow}n}|^2 \rangle$
proportional to the probability ${\cal P}_{n\uparrow}$ of finding ``spin-up,'' 
while the
distribution of $|C_{{\downarrow}n}|$ should also be Gaussian but
with a slightly smaller mean value $\langle
|C_{{\downarrow}n}|^2 \rangle$ proportional to the probability
${\cal P}_{n \downarrow}$ of
finding ``spin-down''. Both ${\cal P}_{n \uparrow}$ and 
${\cal P}_{n \downarrow}$ are very close to ${1 \over 2}$. Thus, taking
into account only the leading order corrections to Eq.(\ref{p_inf}),
we obtain
\begin{equation}
p_n(|C_{\uparrow n}|, |C_{\downarrow n}|)  = 
{\pi \over 2 N_{\hbox{\small b}}}
\hbox{exp} 
\left[ - {N_{\hbox{\small b}}  \over 4} 
\left({|C_{\uparrow n}|^2 \over {\cal P}_{n \uparrow} }+ 
{|C_{\downarrow n}|^2 \over {\cal P}_{n \downarrow} } \right) \right].
\label{p_n}
\end{equation}

We assume  that  in  the course of
the nonequilibrium
evolution, the probability distribution 
$P_n(t, |C_{\uparrow n}|, |C_{\downarrow n}|, \Phi_{n})$
will ultimately evolve to 
\begin{equation}
P_n(\infty, |C_{\uparrow n}|, |C_{\downarrow n}|, \Phi_{n}) = 
{1 \over 2 \pi} p_{\infty}(|C_{\uparrow n}|, |C_{\downarrow n}|),
\label{P_inf}
\end{equation}
where $p_{\infty}$ is given by Eq.(\ref{p_inf}).

Comparing the block trajectories in the quantum case with the
trajectories of the spin vectors on a sphere in the classical case,
three differences can be observed.

The first difference is that, unlike the trajectories 
of individual classical spins, the block trajectories introduced 
for the description of the $n$th spin  describe the
whole system, because the coefficients $C_{\uparrow (\downarrow) n
\alpha}(t)$ are taken from the many-body wave function.

The second difference is closely related to the first one.  Namely,
unlike the case when one sample point representing a given spin moves
on a spherical surface in the course of a sample evolution of the whole
system, many  sample points corresponding to different blocks of the
sample pure state actually move simultaneously in the space $\{
|C_{\uparrow n}|, |C_{\downarrow n}|, \Phi_{n} \}$. 
Moreover, those sample points interact with each other,
because the corresponding
blocks are coupled by the quantum Hamiltonian.
In other words, if we consider a given  block
trajectory constructed for the $n$th spin and
representing the evolution of a pair of coefficients
$C_{\uparrow  n \alpha}(t)$ and $C_{\downarrow n \alpha}(t)$,
then  the immediate environment of that trajectory
is characterized not only by the probability distributions 
$P_k(|C_{\uparrow k}|, |C_{\downarrow k}|, \Phi_{k})$ of
the spins that interact with the $n$th spin, but also by the
distribution $P_n(|C_{\uparrow n}|, |C_{\downarrow n}|, \Phi_{n})$ itself.

Finally, the third difference is that, in comparison with the spherical
surface, which is periodic in all directions and, therefore, finite,
the space of variables $\{ |C_{\uparrow n}|, |C_{\downarrow n}|,
\Phi_{n} \}$ is periodically closed only in the direction of
$\Phi_{n}$.  The situation is less straightforward along the
$|C_{\uparrow n}|$ and $|C_{\downarrow n}|$-directions: Both variables
are bound by zero from below, and also  there is a global normalization
condition $\sum_{\uparrow (\downarrow) \alpha}|C_{\uparrow (\downarrow)
n \alpha}|^2 =1$, which imposes the upper bound $|C_{\uparrow
(\downarrow) n }| \leq 1$.  However, the real upper bound is not $1$
but a much smaller value imposed by the statistical constraint; this
could already be seen from the fact that, according to both the initial
(\ref{p_n}) and the final (\ref{p_inf}) probability distributions , the
typical value of the variables $|C_{\uparrow (\downarrow) n }|$ is of
the order of $1/\sqrt{N_{\hbox{\small b}}}$, where the number of the
basis states $N_{\hbox{\small b}}$ is exponentially greater than
supposedly large number of spins in the system. If evolution of a pure
state starts from the values of all variables $|C_{\uparrow
(\downarrow) n \alpha}|$ of the order of $1/\sqrt{N_{\hbox{\small
b}}}$, then it is improbable that any of those variables can ever
become close to $1$. What is overwhelmingly probable is that any
$|C_{\uparrow (\downarrow) n \alpha}(t)|$ will stay within an upper
bound of the order of $1/\sqrt{N_{\hbox{\small b}}}$ during a time
interval of any reasonable length.

Given all the above differences,
the block trajectories still have two properties
allowing us to treat them in  very much the same way as
in Sections~\ref{cl-intro} and \ref{cl-cont} we treated
the classical trajectories on the spherical surface. These two properties
are:
(a) the mean free time of
the block trajectories  is given by the same one-spin interaction 
time $\tau$ defined  by Eq.(\ref{tau});
and 
(b) the mean free path
of the block trajectories
is of the order of the size of the
finite volume to which those trajectories are constrained,
i.e. 
$\sim \pi$ along the $\Phi_{n}$-direction and 
$\sim 1/\sqrt{N_{\hbox{\small b}}} $ along the 
$|C_{\uparrow (\downarrow) n}|$-directions. 

The reason why
the one-spin interaction
time characterizes the block trajectories is that, 
as was already mentioned, in the ``interaction representation'' we chose
earlier,
the evolution of the coefficients $C_{\uparrow (\downarrow) n \alpha}(t)$
is directly controlled only
by the interaction of the $n$th
spin with its neighbors --- without that interaction,
the coefficients $C_{\uparrow (\downarrow) n \alpha}$ would be
time-independent. 

The mean free
paths given above originate from a simple estimate of 
how much
the value of the individual coefficient $C_{\uparrow (\downarrow) n \alpha}$
can change over an interval of the order of $\tau$ --- subject to the 
condition
that all
coefficients dynamically coupled with a given one have 
absolute values of the order of $1/\sqrt{N_{\hbox{\small b}}}$
and arbitrary
complex phases.

Now we make the hypothesis  that the block trajectories exhibit the
chaotic mixing property  of the same kind as was assumed in
Section~\ref{cl-intro} for the trajectories of classical spins.
Namely, we assume that on the scale characterized by the mean free time
$\tau$: (i) each block trajectory loses the memory of the initial
position; (ii) a set of all block trajectories starting from the
initial positions within an arbitrarily small subvolume of the
statistically constrained part of the space of variables $\{ |C_{\uparrow
n}|, |C_{\downarrow n}|, \Phi_{n} \}$ disperses over  that part in a
random manner; and (iii) the statistics of the trajectory patterns of this
set becomes representative of the statistics of the whole ensemble of
the block trajectories.  In correspondence with the previous
discussion, the term ``statistically constrained'' implies not too
large values of $|C_{\uparrow (\downarrow) n \alpha}|$ in
comparison with $1/\sqrt{N_{\hbox{\small b}}}$.

Most of the arguments we used in 
Section~\ref{markovian} in order to justify {\it Conjecture I}
were given in a general form and apply directly
to the quantum problem posed in terms of the block trajectories.  
The only exception is the phase space
picture motivated by the properties of hyperbolic chaotic systems
(Section~\ref{hyperbolic}).
In principle, one could try to introduce a similar picture in the 
$2 N_{\hbox{\small b}}$-dimensional space
representing the absolute values and the phases
of all coefficients $C_{\uparrow (\downarrow) n \alpha}$, but we do not
pursue this line. Instead, we rely on the general statement
made in Section~\ref{conj_limits} that
{\it Conjecture I} only requires that the environment which governs
the evolution of a given subsystem has a continuous dynamics and a
continuous probability distribution in the space of its variables.
The environment of a given block trajectory 
certainly satisfies the above condition.
(See the preceeding discussion of the ``second difference'' between the
classical spin trajectories and the block trajectories.)

Starting from this point, our entire treatment of the classical spin
systems can be directly translated to the quantum case. Below, we
discuss the quantum case only to the extent that allows us to establish
a correspondence with the mathematical construction obtained in
Section~\ref{cl-cont} for the
classical case.

As in the classical case, we apply {\it Conjecture I} to the ensemble
of the block trajectories and thus justify
Eqs.(\ref{flux},\ref{dfn}), in which $x_n$ now represents the three variables
\{ $|C_{\uparrow n}|$, $|C_{\downarrow n}|$, $\Phi_{n}$ \}; 
and 
\begin{equation}
f_n(t, x_n) = P_n(t, x_n) - P_n(\infty, x_n), 
\label{f_quant}
\end{equation}
where $P_n(\infty, x_n)$ is given by Eqs.(\ref{P_inf},~\ref{p_inf}).

As we have already mentioned, the statistical 
constraint limits the evolution of the block
trajectories to a finite volume.
From the viewpoint of the diffusion description, that constraint implies
that the diffusion
coefficients and the integral kernels in Eq.(\ref{flux}) rapidly
approach zero  as the values of variables 
$|C_{\uparrow (\downarrow)n}|$ become greater than 
$1/\sqrt{N_{\hbox{\small b}}}$.

The solution of the set of Eqs.(\ref{flux},\ref{dfn})
yields the distribution function $f_0(t, x_0)$, from which
the correlation function of interest can be obtained as the
average polarization of the zeroth spin, i.e.
\begin{equation}
G(t) \simeq \langle |C_{\uparrow 0}|^2 \rangle (t) - 
\langle |C_{\downarrow 0}|^2 \rangle (t).
\label{G_quant}
\end{equation}

For any wave
vector {\boldmath $q$} corresponding to the spatial period commensurate
with the lattice periodicity,
the same symmetry argument
as in the classical case 
guarantees that
the diffusion problem (\ref{flux},\ref{dfn}) corresponding to
the calculation of the correlation function
(\ref{G}) can be reduced to Eq.(\ref{diffusion}).

Even though Eq.(\ref{diffusion}) now describes the block trajectories,
the finite volume argument asserting the discreteness of the spectrum
is still applicable, and so are the estimates given in
Section~\ref{cl-cont}. Thus, similarly to Eq.(\ref{average}),
Eq.(\ref{G_quant}) can be expanded in terms of the discrete set
of exponential decays:
\begin{eqnarray}
G(t) &\simeq & 
\int_{|C_{\uparrow 0}|, |C_{\downarrow 0}|, \Phi_0} 
\left (
|C_{\uparrow 0}|^2  - |C_{\downarrow 0}|^2 
\right) \;
f_0(t, |C_{\uparrow 0}|, |C_{\downarrow 0}|, \Phi_0) \;\; 
d|C_{\uparrow 0}| \; d|C_{\downarrow 0}| \; d\Phi_0 
\nonumber 
\\
&=&
\int_{|C_{\uparrow 0}|, |C_{\downarrow 0}|, \Phi_0} 
\left (
|C_{\uparrow 0}|^2  - |C_{\downarrow 0}|^2 
\right) \;
\sum_{\lambda} 
e^{- \lambda t} u_{\lambda}(t, |C_{\uparrow 0}|, |C_{\downarrow 0}|, \Phi_0) \;\; 
d|C_{\uparrow 0}| \; d|C_{\downarrow 0}| \; d\Phi_0 
\nonumber
\\
\label{G_quant1}
\end{eqnarray}
Therefore, we come to the conclusion that the
asymptotic long-time behavior of $G(t)$ has the functional form
(\ref{longexp}) or (\ref{longcos}).  Such a behavior should become
pronounced after the time of the order of several $\tau$.

\subsection{Summary}
\label{summary}

To summarize what was explained by our theory, 
we now give
the answers to the questions posed in the end of Section~\ref{goal}.

Why is the long-time regime universal? 

In the many-dimensional space representing all 
dynamical variables of the problem, the long-time regime is characterized
by a pattern of the probability distribution, which is {\it intrinsic} to a given dynamical 
system (e.g. expansion-contraction pattern) 
and has the property that, on one hand, it is extremely complex,
but, on the other hand, it develops cells of an extremely smooth structure
that spread over the entire many-dimensional space.

What makes the long-time regime different from the initial regime?

The initial probability distribution is factorized, and,
therefore, its pattern is qualitatively
different from the intrinsic one.

Why does the long-time decay have functional form 
(\ref{longexp},\ref{longcos})?

The functional form of the long-time decay is determined
by the slowest eigenmode of the problem  of correlated diffusion in
a finite volume. The finiteness of the volume is the consequence of
two features of our quantity of interest. 
Namely, (i) we deal with spins, which means the finite volume 
of the phase space per lattice site; and (ii) we consider $q$-dependent
correlation functions (see below).

Why are Eqs.(\ref{longexp},\ref{longcos}) 
primarily relevant to  the 
$q$-dependent correlation functions (\ref{G})
and not to the pair correlation functions
of the form $\left\langle \;  S_k^\mu(t)  \; 
S_n^\mu(0) \; \right\rangle$?

In order for the diffusion problem to be posed in 
a finite volume, it is necessary that, at $t=0$, 
there are no more than a finite
number of non-equivalent probability distributions
on different lattice sites. 
The initial probability distributions for  the diffusion problem
associated with a $q$-dependent correlation function
are, actually, all equivalent --- consequence of the fact
that they correspond to the irreducible
representations of the group of lattice translations. 
For the diffusion problem associated with a pair
correlation function, the number of non-equivalent probability
distributions is infinite.
One can only establish the exponential ceiling for the
long-time decay in such a problem. 
It can be done by expanding the pair correlation
function in the infinite series of the $q$-dependent correlation
functions.The ceiling would then be given
the $q$-dependent correlation function having the slowest long-time decay.

What mechanism is responsible for the oscillations in
Eq.(\ref{longcos})? 
 
The resulting diffusion description is non-Hermitian, which
can be linked to the asymmetry in the motion of two interacting spins. 
The eigenvalues of the non-Hermitian diffusion problem
can have both real and imaginary parts. If present, the imaginary part
of the slowest eigenmode leads to the oscillations.

\section{Factors affecting the competition between the monotonic and
 the oscillatory regimes}
\label{factors}

The analysis to be presented in this Section cannot prove by itself that 
the correlation functions (\ref{G}) should have
the long-time behavior either of  form  (\ref{longexp}) or (\ref{longcos}). 
However, once
the functional form of that behavior is established, it is not
difficult to formulate semi-empirical rules that would allow one to
anticipate the outcome of the competition
between the monotonic and the oscillatory regimes.
Such an analysis is possible, because the trends towards the monotonic or
the oscillatory regime can already be extracted from the initial behavior
of the correlation functions, which is, otherwise, not describable
by the functional forms (\ref{longexp}) or (\ref{longcos}). 

To begin with, the long-time
regime of the correlation
functions (\ref{G})  depends on the form of
the Hamiltonian, the value of the wave vector and the choice of the
spin component $\mu$ (i.e. $x$, $y$ or $z$). 
A change in any one of the above three conditions can 
modify the behavior
of the correlation functions from monotonic to oscillatory 
and  vice versa.

We shall exemplify our analysis by
the correlation functions (\ref{G}) involving the $x$-components
of spins (i.e. $\mu \to x$ in Eq.(\ref{G})). This covers
all examples presented in Fig.~\ref{evidence},
with the only exception of Fig.~\ref{evidence}(b)(II).   

We identify the following factors
as important for discriminating between the two regimes:
(i) direct correlations; (ii) motional narrowing;
(iii) indirect correlations; (iv) number of interacting
neighbors; and, finally, (v) quantum-to-classical crossover.
These factors are listed
in the order of their relative importance as derived from our
experience. This order applies to most but not to every
situation.

The factor of direct correlations 
can be evaluated by comparing two quantities. 
The first of them is  the second moment
of $G(t)$:
\begin{equation}
M_2 \equiv  - \left.  {d^2G \over dt^2}\right|_{t=0} 
\label{M2q} 
     =   
{1\over 3} \;   S(S+1) \; \sum_n
\hbox{$ [{J_{kn}^y}^2+{J_{kn}^z}^2 
- 2 J_{kn}^y J_{kn}^z $cos({\boldmath $q \cdot r$}$_{kn}$)$]$};
\end{equation} 
and the second one is $M_{2u}$ --- the second moment of the one-spin 
autocorrelation function 
\begin{equation}
u(t) = {\langle S_k^x(t) S_k^x(0) \rangle \over 
\langle S_k^x(0) S_k^x(0) \rangle }.
\label{u}
\end{equation}
The value of $M_{2u}$ is:
\begin{equation}
M_{2u} \equiv - 
\left.   {d^2u \over dt^2}
\right|_{t=0}=  {1\over 3}
\; S(S+1) \;  \sum_n [{J_{kn}^y}^2 + {J_{kn}^z}^2 ].
\label{M2u}
\end{equation}

The inequality $M_2 < M_{2u}$ indicates that the correlation between
the average polarizations of  neighboring spins  decreases the rate of the
initial decay of $G(t)$ in comparison with $u(t)$.  In this case,
$G(t)$ tends to exhibit  monotonic behavior.  In the extreme case when
$M_2 \ll M_{2u}$, the characteristic time of the decay of $G(t)$  is
much longer than the mean free time $\tau$. This justifies the
Markovian approximation which would predict simple exponential
decay.

In the opposite case, when  $M_2 > M_{2u}$,
the direct correlations favor oscillatory decay. 
It is, however, impossible to propose a Hamiltonian of  form
(\ref{H}) that would lead to $M_2 \gg M_{2u}$. The best one can get
is $M_2 = 2 M_{2u}$, which is the case for Fig.\ref{evidence}(c)(I).

The difference between  $M_2$ and $M_{2u}$ is due to the term 
$2 J_{kn}^y J_{kn}^z $cos({\boldmath $q \cdot r$}$_{kn}$) 
in Eq.(\ref{M2q}).
Therefore, the sign of this difference depends on the sign of
cos({\boldmath $q \cdot r$}$_{kn}$) and on the relative sign of 
$J_{kn}^y$ and $J_{kn}^z$.

The dependence on the relative sign of $J_{kn}^y$ and
$J_{kn}^z$ is, at first sight, difficult to grasp intuitively.
The origin of that dependence, however, becomes 
obvious when Hamiltonian (\ref{H}) is decomposed
in the following way\cite{Fine1}:
\begin{eqnarray}
\nonumber
{\cal H}& = & \sum_{k<n} J_{kn}^x S_k^x S_n^x  
+ {1 \over 4} (J_{kn}^y - J_{kn}^z) ( S_k^+ S_n^+  +  S_k^- S_n^-)  
\\
 & &  
+  \; {1 \over 4} (J_{kn}^y + J_{kn}^z) ( S_k^+ S_n^-  +  S_k^- S_n^+),
\label{H2}
\end{eqnarray}
where $ S_k^+ = S_k^y + i S_k^z $, and $ S_k^- = S_k^y - i S_k^z $.

As far as the correlation functions
of the $x$-components of spins are concerned, 
only the 
second and the third terms in the above decomposition 
are responsible for the
direct correlation effect.  We refer to these two terms as
the ``double-flip term'' and the ``flip-flop term'' respectively.
The correlation trends associated with
them are precisely opposite.
When $J_{kn}^y$ and $J_{kn}^z$ have the same sign,
the flip-flop term,  dominates. In this case, the $x$-components of 
two interacting spins tend to change in the opposite directions,
which, in the extreme case of $J_{kn}^y = J_{kn}^z$, 
leads to the conservation of
the $x$-component of the  total spin. 
When $J_{kn}^y$ and $J_{kn}^z$ have  opposite signs,
the double-flip term dominates, and, as a result the $x$-components of
two interacting spins tend to change simultaneously in the same direction.

Each of the two kinds of direct correlations can 
favor either the monotonic or the oscillatory long-time behavior ---
the correspondence here can  be made definite only after 
the wave vector {\boldmath $q$} is specified.
For example, when  $q=0$, the double-flip term always
favors the oscillatory decay, and the flip-flop term always favors
the monotonic decay.

The direct correlation factor alone would favor  oscillatory decay
in the examples shown in Figs.~\ref{evidence}(a) and (b) and in
Figs.~\ref{evidence}(c)(I) and (III), and  monotonic decay in
Fig.~\ref{evidence}(c)(II). 
It is thus highlighted by 
Fig.~\ref{evidence}(c)(III) that 
there are other important factors to consider.

The second factor among those listed earlier is motional narrowing.
In our definition, motional narrowing is associated with the fact
that, in general, 
the lifetimes   of  local-fields  generated by the Hamiltonian (\ref{H})
are finite. (In other words, those local fields change in time.) 
Here, we are concerned only
with the local field components 
contributing to $M_2$ (i.e. the $y$- and the $z$-components).

The reference to ``narrowing'' originates from the limit of very short
lifetimes. In that limit, a Markovian approximation applies and, as usual,
justifies  simple exponential decay. That decay
is much slower than one would expect just from the
knowledge of $M_2$. 

According to the Hamiltonian (\ref{H}), 
the  local field {\boldmath $h$}$_k$ 
that affects
the $k$th spin is given by 
\begin{equation}
\hbox{{\boldmath $h$}$_k$} = \sum_{n} \left[ J_{kn}^x \; S_n^x  \; 
\hat{\hbox{{\boldmath
$e$}}}_x
+ J_{kn}^y \;  S_n^y \;  \hat{\hbox{{\boldmath $e$}}}_y 
+ J_{kn}^z \;  S_n^z \;  \hat{\hbox{{\boldmath $e$}}}_z \right],
\label{h}
\end{equation}
where $\hat{\hbox{{\boldmath $e$}}}_x$, $\hat{\hbox{{\boldmath $e$}}}_y$
and $\hat{\hbox{{\boldmath $e$}}}_z$ are the unit vectors in the
respective directions.

The lifetime of the $y$-component of the local field can be estimated
as  
\mbox{$[ S^2 \; \sum_n   ({J_{kn}^x}^2 + {J_{kn}^z}^2 )]^{-1/2}$}, 
and the lifetime
of the $z$-component  as 
\mbox{$[S^2 \sum_n  ({J_{kn}^x}^2 + {J_{kn}^y}^2 )]^{-1/2}$}.

Motional narrowing
does not affect the second moments of correlation
functions (\ref{G}), but it does affect the  higher moments.
The effect of motional narrowing is always  to shift
the balance  towards monotonic decay.
The shorter are  the lifetimes of the local fields, the stronger is the
above trend.

The best way to expose the influence of motional narrowing is to
change  $J_x$, because the corresponding term in the Hamiltonian does
not affect either $M_2$ or $M_{2u}$, i.e. it does not contribute
to the direct correlation effect discussed earlier. At the same time,
the presence of that term reduces the lifetimes of both the $y$- and
the $z$-components  of the local-fields.

Let us consider the example shown in Fig.\ref{evidence}(c)(III). 
In this
case, the direct correlation effect is modest. It favors 
oscillatory behavior, which would be the case if $J_{kn}^x = 0$. 
However, the value of $J_{kn}^x$  is sufficiently large. As a result,
the motional narrowing ultimately outweighs the direct
correlations, and the decay becomes monotonic. 

The $J_{kn}^x$-term is also
important for the example shown in Fig.\ref{evidence}(c)(II). 
Although the
direct correlations in this case already favor  monotonic decay,
that trend would be much weaker without the $J_{kn}^x$-term.

The third factor to be discussed here is indirect correlations.
Under this name, we refer to the
correlations between spins that do not interact directly.
These correlations cannot be reflected in
the values of the second moments 
$M_2$ and $M_{2u}$, but, otherwise, 
they are very similar to direct correlations.

We cannot formulate a simple rule for the trend caused by 
indirect correlations.  We can only assert that, in general, 
indirect correlations do not modify the effect of direct
correlations qualitatively.  However, one has to be aware that the
presence of indirect correlations introduces a substantial
uncertainty to the earlier analysis, when the transfer of the 
$x$-polarization
between different spins is efficient (i.e. $|J_{kn}^y| \approx |J_{kn}^z|$), 
but
the wave vector {\boldmath $q$} is such that  the effect of direct
correlations (i.e. the difference between $M_2$ and $M_{2u}$) is
anomalously small.
When such a coincidence occurs (see e.g. fig.1(c) in Ref.~\cite{Fine3}), 
our experience indicates that the effect of indirect correlations
can be anticipated by
assuming, that there exists
an additional interaction of flip-flop type  between 
the next-nearest neighbors.

The rationale for such a prescription is the following. In the leading
order, the 
indirect correlations between two spins are 
due to
their direct interaction with one common neighbor. 
These correlations involve
either two double-flips or two flip-flops, but
in the both cases, the resulting effect is of flip-flop type.
The subtlety here is that nearly the same mechanism
is responsible for
the back-reaction effect, when the polarization of a given spin
is transferred to its neighbor and then transferred back.
Empirically, however,  the indirect correlations coming from the
next-nearest neighbors appear to dominate over the back-reaction
effect.

The fourth factor in our list is the number of
interacting neighbors. From our experience\cite{Fine3}, a smaller 
number of interacting neighbors tends to favor the oscillatory
behavior, though, typically, this effect is not very significant.

Everything in the
preceding analysis  equally applies to both classical and quantum spin
systems.  However, the substitution of classical spins by quantum ones
(especially spins 1/2) is, by itself, a  factor (the fifth one in our list), 
which, to some extent, favors the oscillatory regime.
This factor is most pronounced in the situation with a few interacting
neighbors. It can certainly be linked to the difference
between the solutions of two-spin problems for spins 1/2
and for classical spins.

Now	 we would like to discuss the transition between
the monotonic and the oscillatory long-time regimes. For that purpose,
it is helpful to bring out one of the results of
our treatment in Section~\ref{theory},
namely, that the long-time behavior of the correlation functions (\ref{G})
is, actually, the sum of many exponential terms, some of which can have
the monotonic form (\ref{longexp}) and others can have the
oscillatory  form (\ref{longcos}).

That notion sheds light on the way in which the transition from the
monotonic to the oscillatory decay proceeds as parameters of the
Hamiltonian change. The likely scenario (which is also supported by our
numerical experience) is that, at the transition, the exponential decay
constants of both the slowest oscillating term and the slowest monotonic
term become equal to each other. Therefore, as the transition approaches, 
the long-time behavior of the correlation
functions should be parameterized as the sum of two terms --- one having
form (\ref{longexp}) and the other having form (\ref{longcos}).  

The above description implies that, 
when the previous analysis does not give an
unambiguous prediction, a transitional situation should be suspected, 
which means 
that the correlation function can 
exhibit a superposition of the two regimes during an extended
initial time interval.
Such a competition is, for example, present in Fig.\ref{evidence}(c)(III),
and also in some of the free induction decays reported in Ref.\cite{MG}.

Finally, we would like to mention a useful rigorous result obtained 
in Ref.\cite{Cowan} (Ch.6.3.7) on the basis of the so-called
Widom's theorem. Namely, the inequality $M_4/M_2^2 < 1.5$, where
$M_4$ is the fourth moment of $G(t)$, guarantees that $G(t)$
has at least one zero, which, in turn, strongly suggests (but not guarantees)
that the long-time decay of $G(t)$ has oscillatory form.

\section{Conclusions}

In this work, the treatment of spin systems was centered around {\it
Conjecture~I}, which extended a Brownian-like formalism to 
non-Markovian space-  and timescales.
If true, that conjecture is  important
for non-equilibrium physics in general. Another result of general
interest is a novel
formalism developed for the quantum systems, in order to apply the above
conjecture. A potentially fundamental issue associated
with {\it Conjecture I} is the absence of apparent limits  beyond
which the exact dynamics of systems with infinite number of
particles invalidates that conjecture.

It should be emphasized, that although we presented {\it Conjecture I}
as a natural extension of  physical and mathematical
experience, it was used to address the properties
of  fast 
microscopic  relaxation, which  are generally
considered very
difficult to access, and for which not much  reliable experience
exists.  At present,  our theory based on {\it Conjecture I} can make
only one solid prediction for the fast relaxation, namely, that the
functional form of the generic long-time behavior of the
infinite temperature spin
correlation functions (\ref{G}) is given by Eqs.(\ref{longexp},
\ref{longcos}). The underlying description, however, has a substantial
explanatory power and also allows estimates to be made.

In a separate development, we have also presented a semi-empirical description
of various factors
that discriminate between the monotonic 
long-time regime and
the oscillatory one.

Finally, we would like to mention that 
both the theoretical estimates and 
the empirical evidence show that the knowledge
of the long-time behavior (\ref{longexp}, \ref{longcos}) has a
practical value, related to the fact that 
this behavior becomes pronounced before the
correlation functions (\ref{G}) decay to impractically small values.

\section{Acknowledgements}

The author is grateful to A.~J.~Leggett for numerous discussions,  to
C.~M.~Elliott and to R.~Ramazashvili for helpful comments on the manuscript,
and
to K. Fabricius for sharing some of his unpublished numerical data.
A large part of this work has been done during the author's stay at
the University of Illinois at Urbana-Champaign, where it
has been  supported in part by the MacArthur Chair endowed
by John~D. and 
Catherine~T. MacArthur Foundation at the University of Illinois and  by 
the Drickamer Endowment Fund at the University of Illinois.
This work has also been supported during the author's stay at 
Spinoza Institute, University of Utrecht, by the 
Foundation of Fundamental Research on Matter (FOM), which is sponsored by
the Netherlands Organization for the Advancement of Pure Research (NWO).

\

\appendix

\section{Asymptotic law of the long-time relaxation and the memory function
approach}
\label{critique}

In this Appendix, we discuss the derivation of the long-time behavior
(\ref{longexp}, \ref{longcos}) by Borckmans and Walgraef\cite{BW,BW1}.
(See also Ref.~\cite{Cowan}).
We recall that, even though the technique used by these authors
is more sophisticated than the memory function approach\cite{Tjon,PL}, 
it describes
the long-time behavior of the free induction decay by the
memory function equation
\begin{equation}
{d G(t) \over dt} = - \int_0^t F(t^\prime) G(t-t^\prime) dt^\prime,
\label{memory}
\end{equation}
where $F(t)$ is the memory function.
This equation can be solved by the method of the Laplace transform,
which leads to the following expression:
\begin{equation}
\tilde{G}(s) = {G(0) \over s + \tilde{F}(s)},
\label{laplace}
\end{equation}
where $s$ is the variable of the Laplace transform, 
and $\tilde{G}(s)$ and $\tilde{F}(s)$ are the Laplace transforms of
$G(t)$ and $F(t)$ respectively.

It is not difficult to see that, under certain conditions (e.g.
when $F(t)$ is Gaussian), $\tilde{G}(S)$ has a discrete set of poles, and,
therefore, the pole closest to the imaginary axis controls
the long-time behavior of $G(t)$, which means that the asymptotic 
behavior must
have the functional form  (\ref{longexp}) or 
(\ref{longcos}).  

In essence, the proof of  Borckmans and
Walgraef consisted of exploiting the above fact in
conjunction with several arguments that the shape of $F(t)$ should not
differ much from Gaussian.  

However, a closer examination of the problem reveals the following detail:  
The long-time behavior of $G(t)$ depends crucially on the long-time
behavior of the memory function $F(t)$.  If the long-time decay of
$F(t)$ is slower than exponential, then it is easy to verify by 
direct substitution in Eq.(\ref{memory}) that $G(t)$ cannot decay
exponentially either. Technically, the issue boils down to the fact
that, when the long-time decay of $F(t)$ is slower than exponential,
the Laplace transform of $F(t)$ is likely to have  a branch cut passing
through $s=0$.
Therefore, Eq.(\ref{laplace}) only helps to
reformulate the difficult problem of the long-time behavior of
$G(t)$ in terms of an even less tractable problem of the long-time
behavior of $F(t)$, for which the crude estimates 
of the overall functional shape
are not sufficient.

\end{document}